\documentclass{PoS}
\usepackage{url}
\usepackage{xspace}
\usepackage[numbers]{natbib}

\title{Gamma-ray binaries}

\ShortTitle{Gamma-ray binaries}

\author{\speaker{Maria Chernyakova}\\
        School of Physical Sciences and CfAR, Dublin City University, Dublin 9, Ireland\\
        E-mail: \email{masha.chernyakova@dcu.ie}}
	\author{Denys Malyshev\\
	Institut f{\"u}r Astronomie und Astrophysik T{\"u}bingen, Universit{\"a}t T{\"u}bingen, Sand 1, D-72076 T{\"u}bingen, Germany \\
	}

\abstract{Gamma-ray binaries are a subclass of high-mass binary systems whose energy spectrum peaks at  high energies (HE, E$\gtrsim$100 MeV) and extends to very high energies (VHE, E$\gtrsim$100 GeV) $\gamma$ rays. In this review we summarize  properties of well-studied non-transient $\gamma$-ray binaries as well as briefly discuss poorly known systems and transient systems hosting a microquasar. We discuss also theoretical models that have been used to describe spectral and timing characteristics of considered systems.}

\FullConference{Multifrequency Behaviour of High Energy Cosmic Sources - XIII - MULTIF2019\\
		3-8 June 2019\\
		Palermo, Italy}

\def\lsi{LS~I~+61$^\circ$~303\xspace}
\def\ls{LS~5039\xspace}
\def\psrb{PSR~B1259-63\xspace}
\def\psrj{ PSR J2032+4127 \xspace}
\def\fgl{1FGL~J1018.6-5856\xspace}

\def\lmc{LMC P3\xspace}

\def\grb{$\gamma$-ray binary\xspace}

\def\hess#1#2#3#4{\ifnum#1#2#3#4=0632
          HESS~J0632+057\xspace
       \else
          \ifnum#1#2#3#4=1832
            HESS~J1832-093\xspace
          \else
            HESS~J#1#2#3#4(??)\xspace
       	   \fi
           \fi}

\def\fermi{\textit{Fermi}\xspace}
\def\flat{\textit{Fermi}-LAT\xspace}
\def\swift{\textit{Swift}\xspace}
\def\xmm{\textit{XMM-Newton}\xspace}
\def\cha{\textit{Chandra}\xspace}

\newcommand{\object}[1] {\textit{#1}}

\begin{document}

\section{Introduction}
The population of Galactic X-ray sources  above 2 keV is dominated by the X-ray binaries, see e.g. \cite{Grimm2002}.
A typical X-ray binary contains either a neutron star (NS) or a black hole (BH) which accrets material from a companion star. In the last case due to angular momentum in the system, accreted material does not flow directly onto the compact object,  forming a deferentially rotating disk around the BH known as an accretion disk.
X-ray binaries can be further divided into two different classes, regardless  the  nature  of  the  compact  object,  according  to  the
mass of the companion star: high-mass X-ray binaries (HMXB) and low-mass X-ray binaries (LMXB). The secondary companion of LMXB systems is a low-mass star, which transfers matter by Roche-lobe overflow. HMXBs  comprise a compact object orbiting a massive OB class star. HMXB  systems are strong X-ray emitters via the  accretion  of  matter  from  the  OB  companion.  At the moment 114 HMXBs \cite{Liu2006} and 187 LMXBs \cite{Liu2007} are known. 

Black  hole  X-ray  binaries  (BHXBs)  are  interacting  binary  systems  where  X-rays  are  produced  by  material  accreting from a secondary companion star onto a  BH primary \cite{Shakura1973}. While some material accretes onto the BH, a portion of this inward falling material may also be transferred from the system via an outflow in the form of a relativistic plasma jet or an accretion disk  wind, see e.g. \cite{BHrev2006} for a review. Currently, the known Galactic BHXB population is made up  of  19  dynamically  confirmed  BHs,  and  60  black  hole candidates \cite{BHcat2016}.  The  vast majority of  these Galactic  BHXBs  are  LMXBs. Most of these systems are transient, cycling between periods of quiescence and outburst. This behaviour is associated with changing geometries of mass inflow and outflow, e.g. \cite{BHrev2006}.

At higher energies, however, the situation is drastically different. While current Cherenkov telescopes have detected around  80 Galactic sources \footnote{see TeV Cat catalogue at $http://tevcat2.uchicago.edu/$}, less then 10 binary systems are regularly observed at TeV energies, see  \cite{2013A&ARv..21...64D, grlb_cta_ch19} for a review. These systems are called gamma-ray-loud binaries (GRLB), as the peak of their spectral energy distribution lies at GeV - TeV energy range.

All GRLB systems host compact objects orbiting around massive young star of O or Be spectral type. This allows
to suggest, that the observed $\gamma$-ray emission is produced in the result of interaction of the relativistic outflow from
the compact object with the non-relativistic wind and/or radiation field of the companion massive star. Note that,
neither the nature of the compact object (BH or NS?) nor the geometry (isotropic or anisotropic?)
of relativistic wind from the compact object are known in the most cases. Only in \psrb and \psrj systems the compact object is known to be a young rotation powered pulsar which produces relativistic pulsar wind. In \psrb system the interaction of the pulsar wind with the wind of the Be star leads to the huge GeV flare, during which up to 80\% of the spin-down luminosity is released on average, and even more on a shorter time scales \cite{2011ApJ...736L..11A,chernyakova15, 2018ApJ...863...27J}.

In all other cases the source of the high-energy activity of GRLBs is uncertain. It can be either accretion onto or dissipation of rotation energy of the compact object. In these systems the orbital period is much shorter than in \psrb and \psrj, and the compact object spend most of the time in the dense wind of the companion star. The optical depth of the wind to free-free absorption is big enough to suppress most of the radio emission within the orbit, including the pulsed signal of the rotating NS, \cite{Zdziarski_2010MNRAS_LSI}, making impossible direct detection of the possible pulsar.

GeV observations revealed a few more binaries visible up to few GeV. Among them are Cyg X-1 \cite{CygX1-2016-Zanin,CygX1_2017_Zdz} and Cyg X-3 \cite{2018MNRAS.479.4399Z} -- most probably BH hosting systems. However contrary to the GRLBs described above these systems are transients and seen only during the flares, or, in the case of  Cyg X-1, during the hard state. In addition to this the peak of the spectral energy distribution of these system  happens at much lower energies than in the case of binaries visible at TeV energies. These observations seems to suggest that wind collision can accelerate particles more efficient that the accretion, but more sensitive observations are needed to prove it and understand the reason. 

In our review we aim to expand and update recent reviews on $\gamma$-ray binary systems~\cite{2013A&ARv..21...64D,paredes19,grlb_cta_ch19} with the most recent publications.
%In what follows we review properties of gamma-ray binary systems 
We start with ``classical'' point-like non-transient binaries (Sec.~\ref{sec:classic}), and after that  we continue with new, poorly known systems(Sec.~\ref{sec:poor}), and transient/extended systems hosting microquasars. (Sec.~\ref{sec:microq}). %In our review we follow a recent review on CTA perspectives for gamma-ray binary systems~\cite{grlb_cta_ch19} and the most recent publications on discussed systems.

%??? Include reference to our CTA paper!!!~\cite{grlb_cta_ch19}
\section{Non-transient gamma-ray binaries}
\label{sec:classic}
\subsection{\psrb}
\object{PSR B1259-63}  was first discovered as part of a search for short-period pulsars with the Parkes 64~m telescopes \cite{1992MNRAS.255..401J}, and was the first  radio pulsar (with a 47.76 ms period) discovered in orbit around a massive Be star LS~2883 \cite{1992ApJ...387L..37J} on a highly eccentric 3.4~yr orbit, see e.g. \cite{2014MNRAS.437.3255S} and references therein. 

Radio observations around periastron show an increase and variability in the dispersion measurement of the pulsed signal as the pulsar passes into the stellar wind, see e.g. \cite{2001MNRAS.326..643J}. This is followed by an eclipse of the pulsed signal from $\approx 16$ days before until $\approx 16$ days after periastron, accompanied by the detection of unpulsed radio emission, see e.g. \cite{2005MNRAS.358.1069J} and references therein. These observations suggest that during this time \psrb enters (or closed from the observer by) dense regions of the stellar wind.
%The unpulsed emission shows a double-peak structure with a maximum around the time of the start and end of the pulsar eclipse, although the shape varies between the periastron passages \cite[e.g.][]{2005MNRAS.358.1069J}. The unpulsed emission originates from the extended pulsar wind nebula, which is shown to extend beyond the binary by observations with the Australian Long Baseline Array \cite{2011ApJ...732L..10M}.

%The Be nature of the star is clear from the strong emission lines that are observed from the source, which originate from the out-flowing circumstellar disc \cite{1992ApJ...387L..37J,1994MNRAS.268..430J,2011ApJ...732L..11N}. 
The optical observations further suggest that the Be star disc is  tilted relative to the orbital plane, see e.g. \cite{1998MNRAS.298..997W}, with the pulsar crossing the disc plane twice per orbit. Observations have shown that the circumstellar disc is variable around periastron, with the strength of the H$\alpha$ line increasing until after periastron, as well as changes in the symmetry of the double-peaked He~{\sc i} line \cite{2014MNRAS.439..432C,chernyakova15, 2016MNRAS.455.3674V}. 

In X-ray band the system is characterised by remarkable similarities during different periastron passages, see e.g.~\cite{chernyakova15} and references therein.  Every passage both unpulsed radio and X-ray fluxes peak before and after periastron, at around the same time as the pulsed radio emission becomes/ends to be eclipsed. This behaviour is usually associated with the pulsar's passing through the  circumstellar disk inclined to the orbital plane. Second X-ray peak is about 1.5 times as high as the first one, which is explained by \cite{Chen2019} as a Doppler boosting near the inferior-conjunction.

%After first being detected at X-ray energies with ROSAT \cite{1994ApJ...427..978C}, observations around periastron have shown a remarkable similarity during different periastron passages. X-ray observations folded over multiple epochs show that the X-ray flux peaks before and after periastron, at around the same time as the pulsed radio emission becomes eclipsed \cite[e.g.][and references therein]{chernyakova15}. This is interpreted as being associated with the time the pulsar passes through the plane of the circumstellar disc. 
%Observations around the 2014 periastron passage also revealed that the rate at which the flux decreased after the second maximum ($\approx 20$\,d after periastron) slowed down and plateaued around 30 days after periastron, at the time when the GeV  $\gamma$-ray emission began to increase rapidly. Extended X-ray emission has also been detected around \psrb, with an extended structure flowing away from the binary; this is suggested to be a part of the circumstellar disc that is ejected from the system and begins to become accelerated outwards by the pulsar wind \cite{2011ApJ...730....2P,2015ApJ...806..192P}.

\begin{figure}[t!]
\begin{center}
\includegraphics[width=0.33\linewidth]{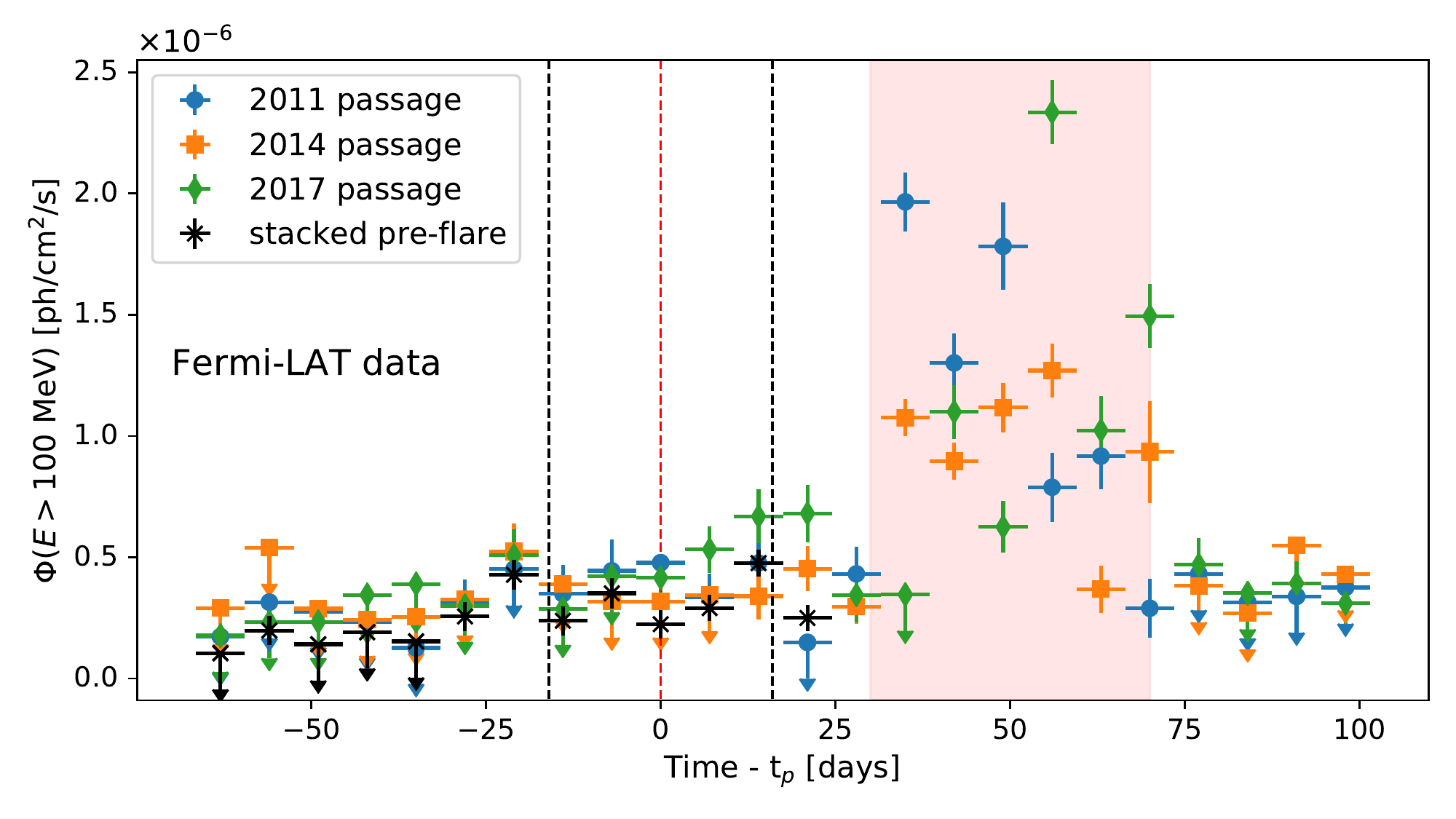}
\includegraphics[width=0.3\linewidth]{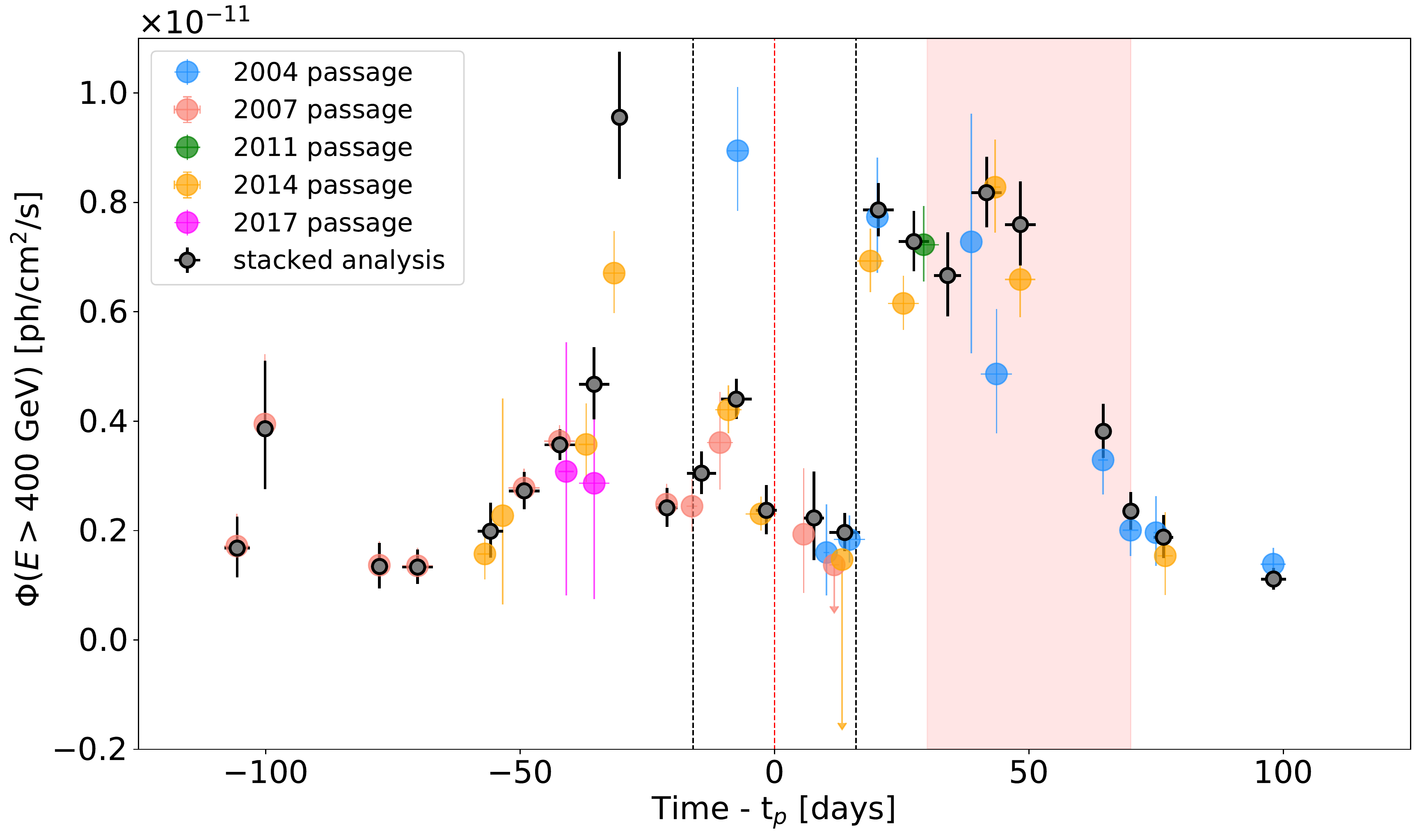}
\includegraphics[width=0.31\linewidth]{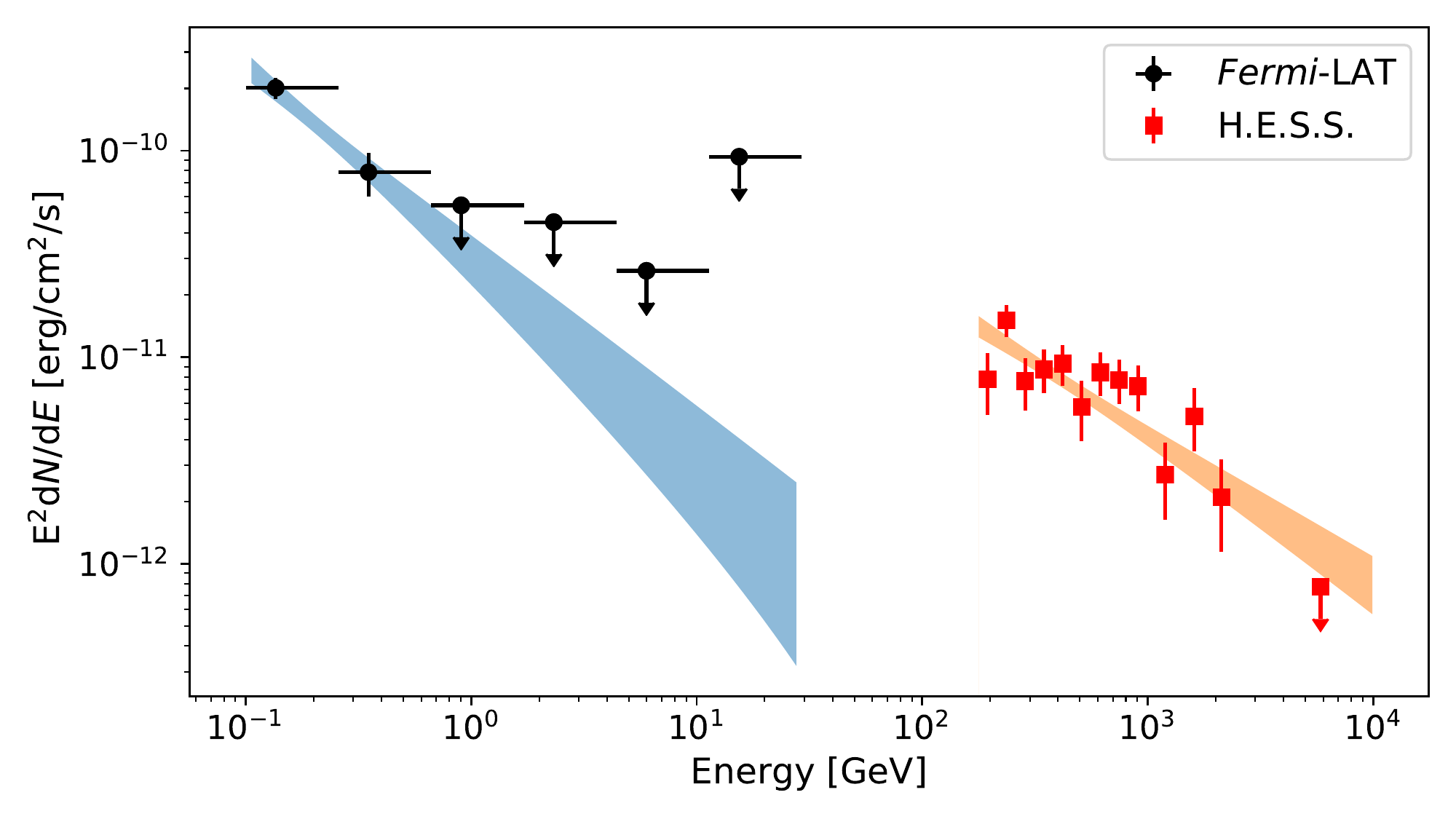}
\end{center}
\vspace*{-0.75cm}
\caption{ Left and middle panels show GeV and TeV lightcurves of \psrb as seen by \flat and H.E.S.S. correspondingly. Right panel present joint \flat/H.E.S.S. SED of the source. All panels are adapted from~\cite{psrb_hess19} }
\label{fig:psrb_fermi_hess}
\end{figure}

In GeV and TeV energies the system has been clearly detected by \flat and H.E.S.S. only at orbital phases close to periastron, see Fig.~\ref{fig:psrb_fermi_hess} and~\cite{psrb_hess19} for a review of observations. While at TeV energies combined over different epochs observations suggest similar to X-rays double-peak lightcurve, the GeV observations reveal very different behaviour. In this band the system is barely detected prior and around periastron, while around $\sim 30-70$ days after periastron, a rapid brightening (flare) with an average luminosity approaching that of the pulsar spin-down luminosity is observed \cite{2011ApJ...736L..10T,2011ApJ...736L..11A,chernyakova15}. While the flare seems to be present after every observed periastron passage, it demonstrate different short-term variability pattern and energetics. During 2017 GeV flare \psrb shown variability on $10$~minutes -- few hours timescales \cite{2018arXiv180409861T,2018ApJ...863...27J}. At the shortest scales the released in GeV band energy significantly (by a factor of $5-30$) exceeds  the total spin-down luminosity.

Several models have been proposed so far to explain the observed multiwavelength emission from this system.
The broadband emission as originating from electrons accelerated in the shock between the pulsar wind and stellar wind and produced by synchrotron and/or IC radiation was suggested by~\cite{tavani_arons, kirk99, Dubus06, bogovalov08, khan11,2014MNRAS.439..432C,Chen2019}. \cite{2017ApJ...837..175S} proposed to explain the periastron dip in the TeV lightcurve to be due to strong gamma-gamma absorption due to stellar and disk photons. 

\cite{chernyakova15} basing on the coincidence of start of the rapid decay of the $H\alpha$ equivalent width with the start of GeV flare argued that the last one is associated to the disruption of the circumstellar disk. The GeV flare was explained as a result of synchrotron cooling of monoenergetic relativistic electrons injected into the system during this event.
The comptonization of unshocked pulsar wind particles~\cite{khan12}  as well as Doppler boosting~\cite{dubus_cerutti10, kong12} were also suggested to play a role in producing the GeV flares. \cite{2017ApJ...844..114Y}  proposed that GeV flare can be a result of transition between the ejector, propeller, and accretor phases. In this model compact object is  working as an ejector all along its orbit and being powered by the propeller effect when it is close to the orbit periastron, in a so-called flip-flop state. However none of the models proposed so far can explain all the details of the GeV flare during the 2014 periastron passage.

%%%%%%%%%%%%%%%%%%%%%%%%%%%%%%%%%%%%%
%%%%%%%%%%%%%%%%%%%%%%%%%%%%%%%%%%%%%

\subsection{PSR J2032+4127}

Shortly after \flat discovery of \psrj as a $\sim 143$~ms gamma-ray pulsar \cite{fermipulsars09}, it was also detected to be a pulsar at radio frequencies  \cite{Camilo09}. The source is located close to the extended TeV HEGRA source, TeV J2032+4130  \cite{hegra05}, and it was proposed that this was a pulsar wind nebula powered by \psrj \cite{aliu14}. However, while \psrj was first thought to be an isolated pulsar \cite{Camilo09}, 
further radio observations demonstrated a rapid increase in its observed spin-down rate, which was interpreted as an evidence that the pulsar is a member of a highly-eccentric binary system, where the optical companion is a $\sim$ 15 M$_{Sun}$ Be star, MT91~213 \cite{lyne15}. \psrj thus turned out to be similar to \psrb, described above.

Further multi-wavelength monitoring of \psrj by \cite{ho17} refined the orbital period  to be in the range of $16\,000$ to $17\,670$ days, with an eccentricity of $e = 0.961$ (separation at periastron is $\alpha_{\rm per}\sim 1$\,au)  and found the  periastron passage would occur in November 2017. Subsequent observations around this period were in good agreement with this prediction.  

\begin{figure}[t!]
\begin{center}
\includegraphics[width=0.26\linewidth]{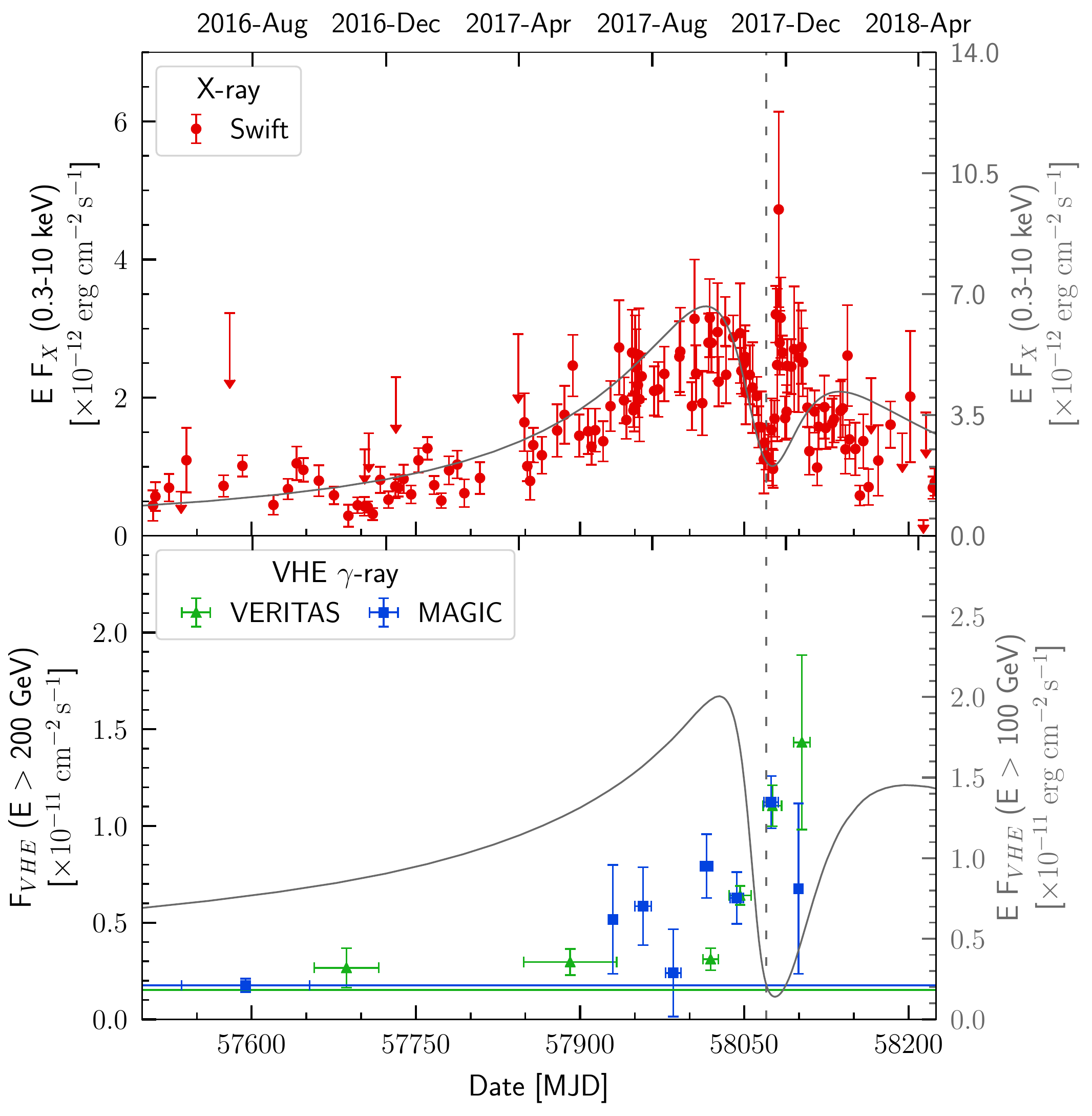}
\includegraphics[width=0.36\linewidth]{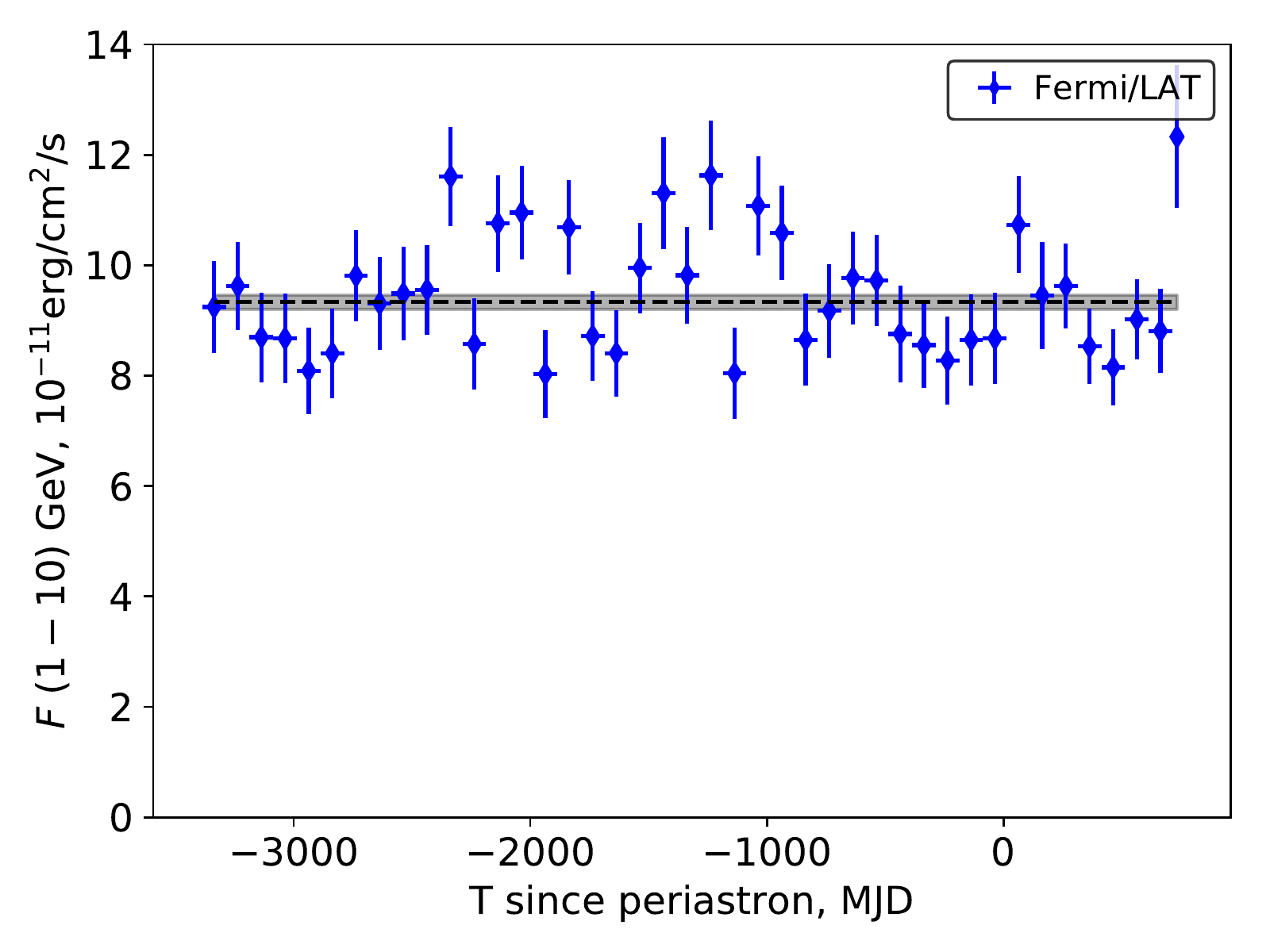}
\includegraphics[width=0.36\linewidth]{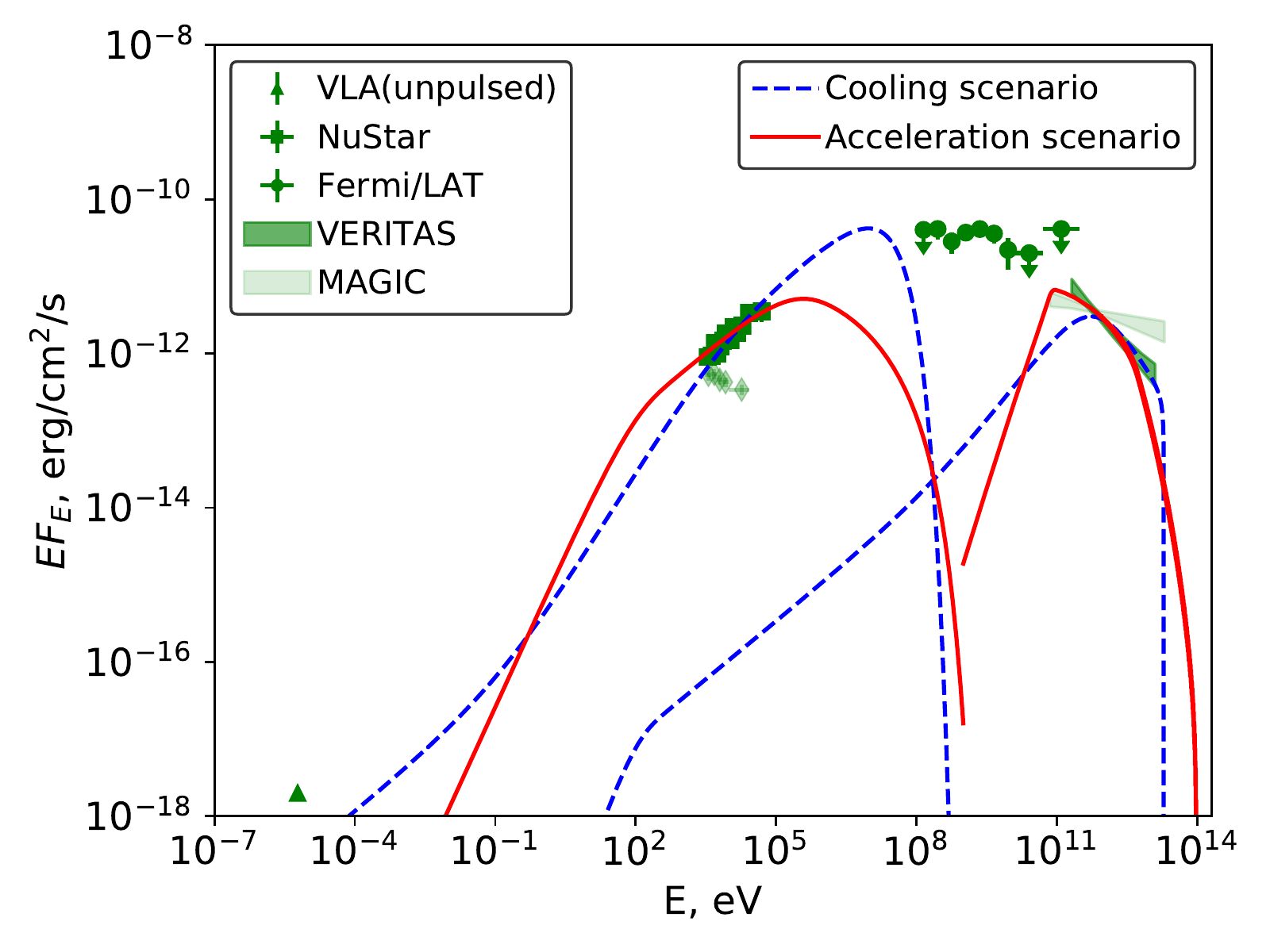}
\end{center}
\vspace*{-0.75cm}
\caption{ GeV-TeV lightcurves and SED of \psrj as seen by \flat and H.E.S.S., adapted from~\cite{psrj_magic} and~\cite{we_psrj2019}. }
\label{fig:psrj_xray_fermi_hess}
\end{figure}

Long-term X-ray observations of \psrj prior to periastron (2004-2016) indicated that the keV-band flux increased by a factor of 70~\cite{ho17}. Close to the periastron X-ray lightcurve of this system resembles one of \psrb with two-peak structure, see Fig.~\ref{fig:psrj_xray_fermi_hess}. The slope of the X-ray band spectrum hardens as the pulsar approaches the Be star, starting from $\sim -2$ far from the periastron and reaching a value of $\Gamma\sim1.5$ around the periastron passage~\cite{we_psrj2019}.

In the TeV band the behaviour of the system is characterised by a complex variation of the flux at around the periastron, see Fig.~\ref{fig:psrj_xray_fermi_hess}, left panel, adopted from~\cite{psrj_magic}. However, contrary to \psrb system, in the GeV band the flux of \psrj remained roughly constant during the whole periastron passage.

The time evolution of \psrj was modelled by~\cite{li18} (radio, X-ray and GeV bands), \cite{coe19}(optics and X-ray) and~\cite{we_psrj2019} (X-ray to TeV band and spectral modelling).  
\cite{li18} suggested that the strong X-ray dip close to periastron is explained by an increase of the magnetization parameter of the star--pulsar colliding winds shock accompanied by flux suppression due to Doppler boosting effect. The
post-periastron rise could be a consequence of the Be stellar disc passage by the pulsar. The absence of variability in the GeV emission was explained by the strong dominance of the pulsar magnetospheric emission over the expected orbital-modulated IC emission. The model was able to predict the overall shape of the orbital X-ray lightcurve, but was not able to reproduce the details of the double-peak flux structure around periastron.% The origin of the hardening of the X-ray spectrum, as the pulsar approaches periastron, also does not appear in the model in a natural way and was attributed by the authors to a possible increase of the hydrogen column density.
%  $N_H$.

\cite{coe19} reported on optical and X-ray flux measurements of \psrj,  accompanied by  SPH modelling of the Be star/pulsar interaction. In their model authors explicitly assumed that the disc of the Be star is inclined to the orbital plane. The modelling, however, failed to describe the details of the observed X-ray lightcurve of the system, generally predicting a maximum of the flux at periastron and has no clear prediction on the variability of the X-ray spectrum. 

Additionally, \cite{ng19} recently presented radio to X-ray observations of \psrj. The authors found that radio and X-ray components can not be connected by a simple power law and thus may originate from different spectral components. The authors also noticed a hint of a spectral break at $\sim 5$~keV energies which was attributed to the modification of the spectrum of accelerated electrons by synchrotron losses.

\cite{we_psrj2019} qualitatively explained observed X-ray to TeV lightcurve of \psrj as well as corresponding spectrum of the source in term of the model which also suggest the inclined to the orbital plane circumstellar disk. The authors argued that the observed spectrum can be formed in terms of either ``acceleration'' (monoenergetic electrons of pulsar wind are additionally accelerating on the shock) or ``cooling'' (monoenergetic electrons are cooled down due to synchrotron and Inverse Compton (IC) losses in the system). Discussing similarities with other inclined-disk gamma-ray binaries the authors argued for a preference of acceleration scenario.
%%%%%%%%%%%%%%%%%%%%%%%%%%%%%%%%%%%%%
%%%%%%%%%%%%%%%%%%%%%%%%%%%%%%%%%%%%%

%\textit{Chandra} X-ray observations reported in \cite{ho17} demonstrated a strong rise in the  source brightness;  \psrj was about 20 times brighter in 2016 than it was in 2010, and about 70 times brighter than it was in 2004. This increase was interpreted as a result of the collision between pulsar and Be star winds. Smoothed-particle hydrodynamics (SPH) modelling of the winds interaction presented in \cite{coe19} demonstrates the deformation and even partial destruction of the Be star disc at some orbital phases.

%Modelling of the source suggested that around periastron the binary should be detectable at TeV energies (\cite{bednarek18,takata17}) and \psrj was subsequently detected at TeV gamma-ray energies by the VERITAS and MAGIC telescopes as a point-like source (\cite{psrj_magic}). The TeV light curve peaked at periastron (followed by a short dip a few days later), see Fig.~\ref{fig:psrj_xray_fermi_hess}. The observations also showed that during the brightest period (at periastron) the spectrum is best fit by a power-law but when the source was fainter during the 2017 observations (before periastron) a power-law with an exponential cut-off is favoured. 

%Long term observations of the Be optical companion, MT91 213, have shown the star has demonstrated strong variability showing periods of marked changes in the emission lines. 

%???What do we write here about our data and model???

\subsection{ LS I +61$^\circ$ 303}
\label{sec:lsi}
\object{LS I +61$^\circ$ 303}
was first discovered as a bright  $\gamma$-ray source by the Cos B satellite \cite{1977Natur.269..494H}. Shortly after the discovery, it was realised that this source was also a highly variable radio source  \cite{1978Natur.272..704G} and was associated with the optical source \lsi, a young, rapidly rotating, 10–15 M$_\odot$ B0 Ve star \cite{1979AJ.....84.1030G}. A young pulsar was at first suggested to be responsible for the observed radio emission \cite{Maraschi1981}, but no pulsations have ever been detected, despite intensive searches (e.g. \cite{2006A&A...459..901S}). Still hidden pulsar is a plausible option, as  the optical depth of the wind due to free-free absorption is big enough to suppress most of the radio emission within the orbit, including the pulsed signal of the rotating NS \cite{Zdziarski_2010MNRAS_LSI}, impeding a direct detection of the possible pulsar.

Radial velocity measurements of the absorption lines
of the primary \cite{2005MNRAS.360.1105C,2009ApJ...698..514A} showed that \lsi is on an elliptical ($e=0.537\pm0.34$) orbit. The orbital period of \lsi\ was found to be $P\approx26.5$~d from radio observations \cite{2002ApJ...575..427G}.
 A strong orbital modulation in \lsi is also observed in the optical to infrared \cite{Mendelson1989,Paredes1994},  X-ray \cite{Paredes1997}, hard X-ray \cite{Zhang2010}, and  HE/VHE $\gamma$-ray \cite{Abdo2009, Albert2009} domains.
In the optical band, the orbital period signature is evident not only in the %visible 
broad-band photometry, but also in the spectral properties of the H$_\alpha$ emission line \cite{Zamanov1999,fortuny15_LSI}. Because of the uncertainty in the inclination of the system, the nature of the compact object remains unclear, and it can be either a NS  or a stellar-mass BH \cite{2005MNRAS.360.1105C}.

In radio, \lsi\ was intensively monitored at GHz frequencies for many years, see e.g. \cite{1997ApJ...491..381R,2015A&A...575L...9M}. The radio light curve displays periodic outbursts whose position and amplitude changed from one orbit to the next. A Bayesian analysis of radio data allowed \cite{2002ApJ...575..427G} to establish a super-orbital periodic modulation of the phase and amplitude of these outbursts with a period of $P_{\rm so}=1667\pm 8$ days.
%, which confirm the initial findings of \cite{1987PhDT.......113P}.
This modulation has also been observed in X-rays \cite{2012ApJ...747L..29C,2014ApJ...785L..19L} and $\gamma$~-rays \cite{2013ApJ...773L..35A,Ahnen:2016,Xing2017}.
%,2018MNRAS.478..440J
It has been suggested that the super-orbital periodicity can depend  on the Be star disc, either due to a non-axisymmetric structure rotating with a %periodicity 
period of 1667 days \cite{Xing2017},  or because of a quasi-cyclic build-up and decay of the Be decretion disc \cite{Negueruela-2001,2013ApJ...773L..35A, 2012ApJ...747L..29C, Chernyakova2017}. Another possible scenario for the super-orbital modulation is related to the precession of the Be star disc \cite{Saha_2016} or  periodic Doppler-boosting effects of a precessing jet \cite{2016AA...585A.123M}.

The precessing jet model is based on high-resolution radio observations suggesting a double-sided jet \cite{Massi1993, Paredes1998,Massi2004}. 
The precession period in this model is about 26.9 days, which is very close to the orbital period.  In this case the observed super-orbital variability is explained as a beat period of the orbital and precession periods \cite{2013A&A...554A.105M}. 

To test the nature of the compact source in \lsi \cite{LSI_2017_BH}  studied the correlation between the X-ray luminosity and the X-ray spectral slope in \lsi and found a good agreement with that of moderate-luminosity BHs. Along with the  presence of 55 minutes and 2 hours long quasi-periodic oscillations in radio and X-rays correspondingly stable over the few days, see e.g. \cite{LSI_QPO_2018}, this supports a microquasar scenario for \lsi.
 However, in this case, it is the only known microquasar that exhibits a regular behaviour, does not demonstrate transitions between various spectral states, and lacks a spectral break up to hard  $\gamma$-rays.  
 
 At the same time, in \cite{Zdziarski_2010MNRAS_LSI} it was shown, that the model in which the compact source is a pulsar allowed
naturally explain the keV-TeV spectrum of LSI +61 303. Authors argued, that the radio source has a complex, varying
morphology, and the jet emission is unlikely to dominate the spectrum through the whole orbit. Within this model
the superorbital period of the source is explained as timescale of the gradual build-up and decay of the disk of the Be star. In \cite{2012ApJ...747L..29C} authors demonstrate presence of the superorbital variability in X-rays and show that a  constant  time  delay  between  the  drifting  orbital phases of X-ray and radio flares could be naturally explained  if  one  takes  into  account  the time needed for electrons to reach  regions transparent for radio emission.

  Cyclical  variations  in  the  mass-loss of  the  Be  star is supported by the optical observations confirming the superorbital variability of the Be-star disk \cite{fortuny15_LSI}, and is also a reason of superorbital variability in an alternative  flip-flop model  \cite{zamanov01,Torres2012,Papitto:2012,Ahnen:2016}. This model assumes compact object in  \lsi to be a magnetar and  implies a change from a propeller regime at periastron to an ejector regime at apastron. During the periastron  the pressure of  matter from the Be star outflow compress and disrupt the  magnetosphere of NS, which lead to the disappearing of the  pulsar wind. In this case electrons are accelerated at the propeller shock, which accelerates electrons to lower energies than  inter-wind shock produced by the interaction of a rotationally powered pulsar and the stellar wind of the Be star. 
A magnetar-like short burst caught from the source supports the identification of the compact object in \lsi\ with a NS \cite{Barthelmy2008, Burrows12, Torres2012,Barthelmy2019}. However, RXTE observations of \lsi demonstrated presence of few  several second  long flares \cite{2009ApJ...693.1621S}, which were compared by  authors to the flares typically found in the accretion  driven  sources. Note however, that neither BAT, nor  RXTE  cannot exclude the possibility that the observed flares are coming from another source located close to the  line-of-sight.

%%%%%%%%%%%%%%%%%%%%%%%%%%%%%%%%%%%%%
\begin{figure}[t!]
\vspace*{-0.5cm}
\begin{center}
\includegraphics[width=0.27\linewidth]{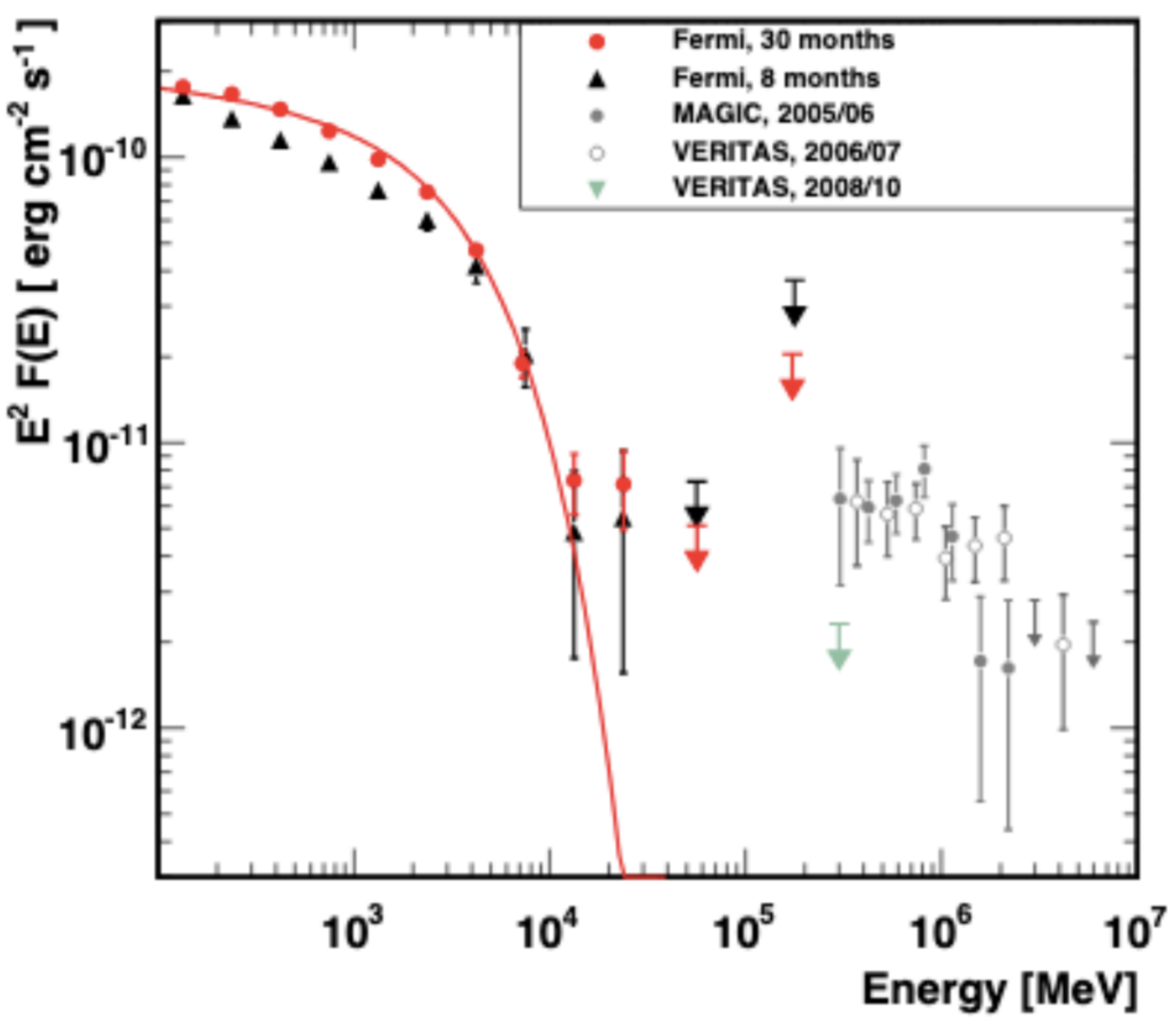}
\includegraphics[width=0.34\linewidth]{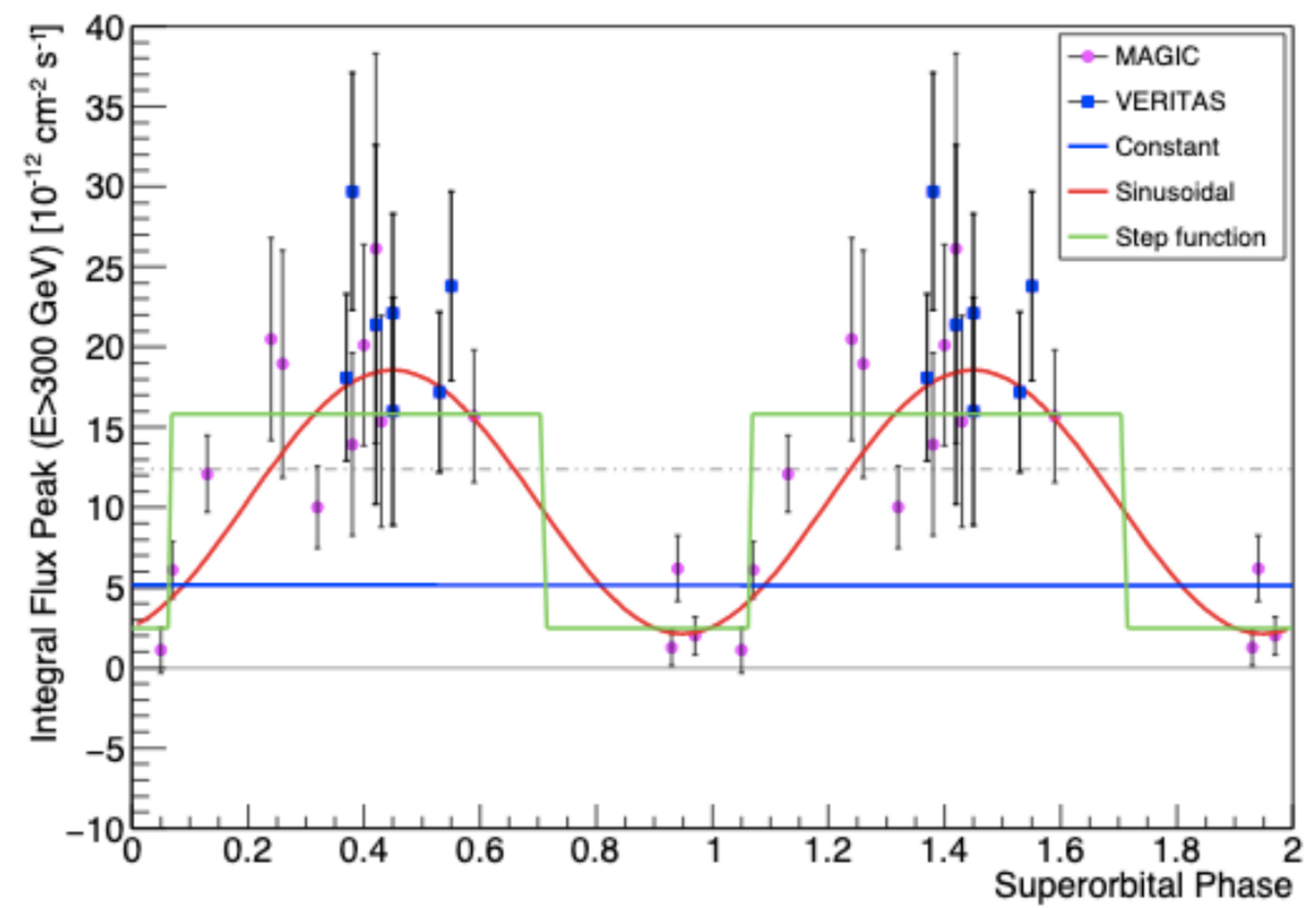}
\includegraphics[width=0.34\linewidth]{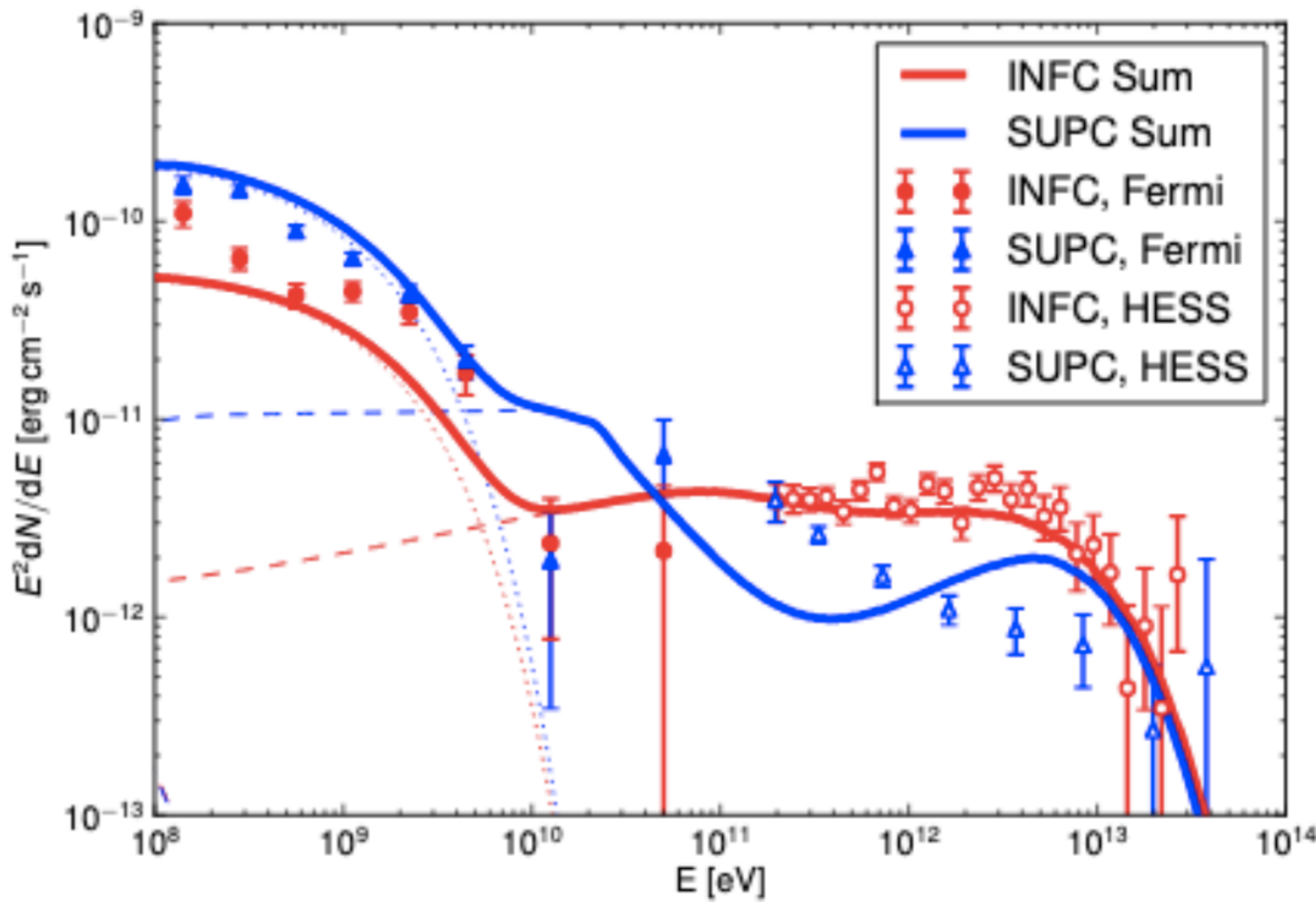}
\end{center}
\vspace*{-0.5cm}
\caption{ \textit{left:}  GeV-TeV  SED of \lsi as seen by \flat, Magic and Veritas, adapted from~\cite{Hadasch2012}.  \textit{middle:} Peaks of the VHE emission in terms of the super-orbital phase, adapted from \cite{Ahnen:2016}. \textit{right} SED of LS 5039. H.E.S.S. data during inferior (red) and superior conjunctions are compared with the model of \cite{Zabalza13}. The emission components from the wind standoff and Coriolis turnover locations are indicated with a dotted and dashed line, respectively. Adapted from \cite{Zabalza13}.
} 
\label{fig:lsi_ls}
\end{figure}
%%%%%%%%%%%%%%%%%%%%%%%%%%%%%%%%%%%%%

At GeV energies, \lsi was unambiguously detected %at high significance 
 by \textit{Fermi}-LAT \cite{Abdo2009} through its flux modulation at the orbital period. 
%The light curve from \flat data shows a broader peak after periastron and a smaller peak just before apastron \cite{2014A&A...572A.105J}.
The \flat light curve shows a broader peak after periastron and a smaller peak just before apastron \cite{2014A&A...572A.105J}.
The peak at apastron is affected by the same orbital shift as the radio outbursts and varies on the super-orbital timescale, leading to a decline in the orbital flux modulation as the two peaks merge. 

A long-term investigation of \textit{Fermi}-LAT data 
by \cite{Saha_2016} showed  the orbital spectral variability of the  source. The observed spectrum is consistent with an exponential cut-off power law with a cut-off at 6--30 GeV for different orbital states of the system. The excess above the spectral cut-off is part of a second emission component that is dominant at the TeV domain \cite{Hadasch2012,Saha_2016}, see right panel of Figure \ref{fig:lsi_ls}.

Detected at TeV energies by MAGIC \cite{Albert2006} and VERITAS \cite{Acciari2008}, the VHE emission from \lsi\ 
shows a modulation consistent with the orbital
period \cite{Albert2009} with the flux peaking at apastron. A decade-long VERITAS observation of \lsi allowed TeV emission to be detected from the system throughout the entire orbit, with the integral flux above 300 GeV varying in the range $(3 - 7) \times 10^{-12} \mathrm{cm}^{-2} \mathrm{s}^{-1}$.
The VHE emission  is well described by a simple power-law spectrum, with a photon index of
$\Gamma=2.63\pm0.06$ near apastron and $\Gamma=2.81\pm0.16$ near periastron \cite{2017ICRC...35..712K}. 

Long-term multi-wavelength monitoring of \lsi indicates a correlation between the X-ray ({\it XMM-Newton} and {\it Swift}/XRT) and TeV (MAGIC and VERITAS) data sets. 
At the same time, GeV emission shows no correlation with the TeV emission. Along with the spectral cut-off at GeV energies, this implies that the GeV
and TeV gamma rays originate from different particle populations   \cite{Anderhub2009,2013ApJ...779...88A,2015ICRC...34..818K}. 

\cite{Bednarek2011} propose that these populations are naturally produced by  electrons accelerated on a double shock structure  created within the binary system as a result of the interaction of the pulsar and massive star winds. The shock from the side of the pulsar is able to accelerate electrons to higher energies than the one from the  side of massive star. These two populations of electrons produce two component $\gamma$-ray spectra caused by the IC scattering of stellar radiation.

Similar to other wavelengths, the TeV %orbital 
light curve varies from orbit to orbit. 
MAGIC observations spanning over  two super-orbital periods \cite{Ahnen:2016} demonstrate that the TeV flux of the periodical outburst around apastron show yearly variability consistent with the long-term modulation of about 4.5 years found in the radio band, see middle panel of Figure \ref{fig:lsi_ls}. There is no evidence for a correlation between the TeV emission and the mass-loss rate of the Be star, but this may be affected by the strong, short-timescale (as short as intra-day) variation displayed by the H$_\alpha$ fluxes.

\subsection{\ls \label{section:ls}}
\object{LS 5039}
has the shortest orbital period thus far of all known  $\gamma$-ray binaries (3.9\,d). It is also known as V497 Sct, based on {\it ROSAT} X-ray data. \cite{1997A&A...323..853M} first reported it as a high-mass X-ray binary. Its peculiar nature as a persistent non-thermal radio emitter was  soon revealed after the detection of a bright radio counterpart with the Very Large Array (VLA) by \cite{1998A&A...338L..71M}. This has anticipated the capability of the system to accelerate 
%%relativistic electrons.
electrons to relativistic energies. Follow-up images obtained with very long baseline interferometry (VLBI) resolved the radio emission into elongated features, 
%%and then \ls\ was interpreted as new microquasar system \cite{2000Sci...288.2340P}.
and as a result, \ls\ was interpreted as a new microquasar system \cite{2000Sci...288.2340P}. Moreover, at the same time, it was also tentatively associated with the EGRET $\gamma$-ray source 3EG J1824$-$1514. The confirmation of \ls\ as an unambiguous ($>100$~GeV) $\gamma$-ray source was finally obtained with H.E.S.S. \cite{2005Sci...309..746A}.

During the 20~yr since its discovery, the physical picture of \ls\ has generally evolved from the microquasar scenario to a binary system hosting a young non-accreting NS  interacting with the wind of a massive O-type stellar companion (see e.g. \cite{2013A&ARv..21...64D} and references therein). This is strongly supported by VLBI observations of periodic changes in the radio morphology \cite{2012A&A...548A.103M}, although no radio pulsations have been reported so far.

At different energies, the shape of the \ls light curve varies, as confirmed in the most recent multi-wavelength studies using {\it Suzaku}, {\it INTEGRAL}, {\it COMPTEL}, {\it Fermi}-LAT, and H.E.S.S. data, see e.g. \cite{2016MNRAS.463..495C} and references therein. The X-ray, soft $\gamma$-ray (up to 70 MeV), and TeV emission peak around 
%the system 
inferior conjunction after the apastron passage. In contrast, $\gamma$-rays in the 0.1-3 GeV energy range anti-correlate and have a peak near the superior conjunction soon after the periastron passage. No clear orbital modulation is apparent in the 3-20 GeV band. GeV-TeV spectra of \ls during superior and inferior conjunctions are shown in the left panel of Figure \ref{fig:lsi_ls}.  

This dichotomy suggests a highly relativistic particle population that accounts for both X-ray/soft $\gamma$-ray and TeV emission mainly by synchrotron and anisotropic IC scattering of stellar photons, respectively. The GeV $\gamma$-ray peak would arise when TeV photons (of an IC origin) are absorbed  through pair production as the NS approaches its O-type companion, and further enhances the GeV emission through cascading effects. Variable adiabatic cooling and Doppler boosting are other effects proposed to play an important role when trying to understand the multi-wavelength modulation of systems such as \ls\ see e.g.~\cite{2008IJMPD..17.1909K, Suzaku2009, 2013A&ARv..21...64D}. \cite{Zabalza13} proposed that GeV and TeV photons are produced  by particle populations originating from the electrons accelerated correspondingly at the shocks formed at the head-on collision of the winds and the termination shock caused by Coriolis forces on scales larger than the binary separation, see left panel of Figure \ref{fig:lsi_ls}.  The lower magnetic field and larger distance to the star, resulting in lower synchrotron and IC losses, favour the acceleration of particles up to very high energies at the at the Coriolis turnover region. 

\subsection{\fgl}

%\fgl\ 
\object{1FGL J1018.6-5856}
(also known as 3FGL J1018.9-5856, HESS J1018-589A) is a first gamma-ray binary identified through blind search for periodic source in \flat data~\cite{LAT2012}. $\gamma$-ray flux from the source was found to be modulated with a period of 16.544~days. The radio and X-ray fluxes are modulated with the same period which was interpreted as a binary orbital period~\cite{LAT2012}.
%It is a point like source positionally coincident with the supernova remnant SNR G284.3–1.8. 

Using the Gaia DR2 source parallax and assuming a Gaussian probability distribution for the parallax measurement, \cite{Marcote+18} derived a source distance of $d = 6.4^{+1.7}_{-0.7}$ kpc. They also calculated the Galactic proper motion of the source and found that it is moving away from the Galactic plane. Both the source distance and proper motion are not compatible with the position of the SNR G284.3-1.8 (which is located at an estimated distance of $\simeq 2.9$~kpc). Therefore, it is possible to exclude any physical relation between the binary source and the SNR.

Optical observations show that the source is positionally coincident with a massive star of spectral type O6V(f). 
Spectroscopic observations of the optical counterpart allowed  \cite{Strader+15}  to find that a companion star has a low radial velocity semi-amplitude of 11-12~km s$^{-1}$, which favours a NS as a compact object. This conclusion is in agreement with the results of \cite{Monageng+17}, who constrained the eccentricity of the orbit $e = 0.31 \pm 0.16$ and showed that the compact object is a NS, unless the system has a low inclination $i \lesssim 26^\circ$. 

TeV observations of 1FGL J1018.6-5856 reveal two TeV sources (point-like HESS J1018-589A and $\sim 0.15^\circ$-extended HESS J1018-589B) within \flat localization of 1FGL J1018.6-5856~\cite{2015A&A...577A.131H}. The flux of HESS J1018-589A source was found to be variable on night-to-night basis and when convolved with the GeV period~\cite{2015A&A...577A.131H}.

\begin{figure}[t!]
\begin{center}
\includegraphics[width=0.26\linewidth]{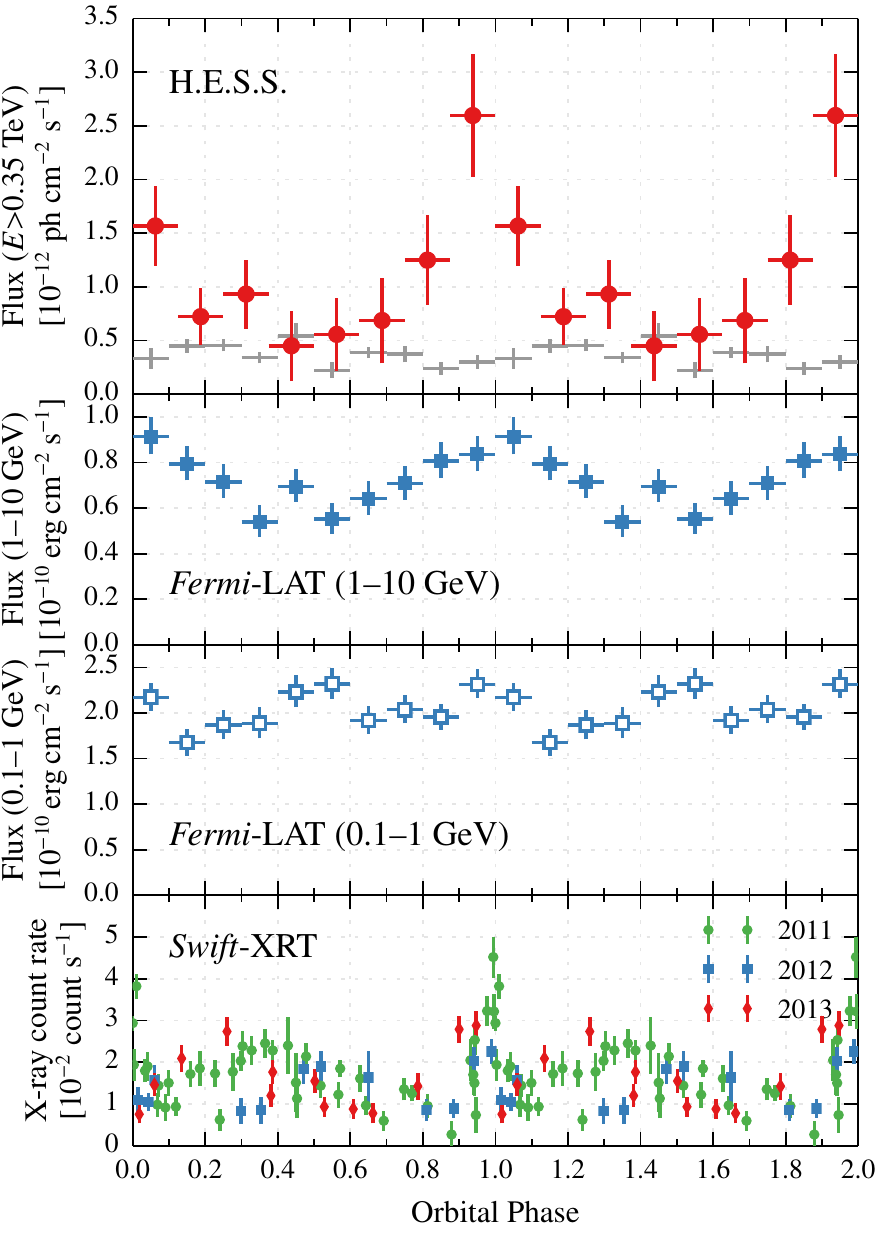}
\includegraphics[width=0.5\linewidth]{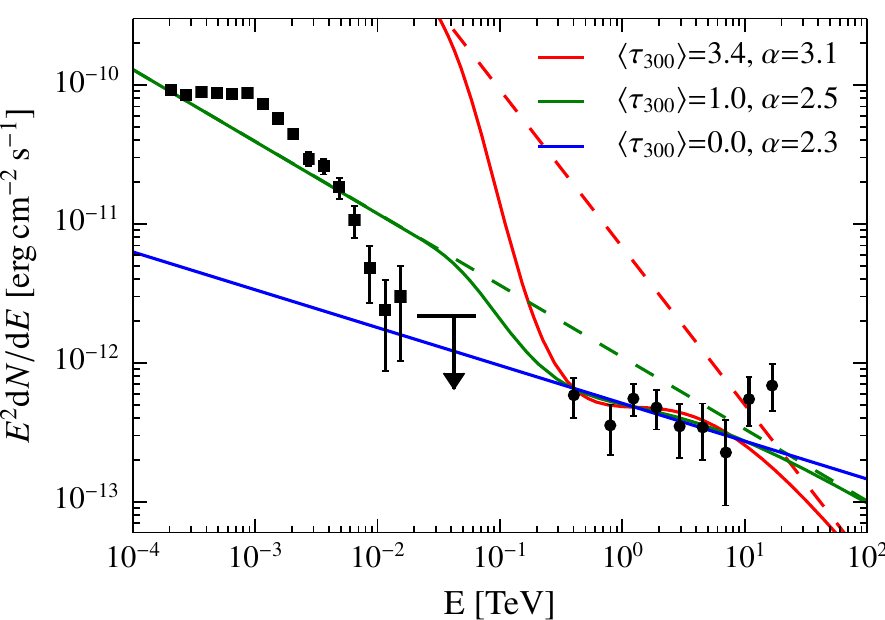}
\end{center}
\caption{ Left: X-ray to TeV lightcurves of \fgl (adapted from~\cite{2015A&A...577A.131H}) Right: GeV to TeV spectrum of \fgl. The $\gamma$-ray absorption effects due to pair production losses are shown with lines (adapted from~\cite{2015A&A...577A.131H}). }
\label{fig:1fgl}
\end{figure}

In a dedicated observation campaign at VHE, HESS J1018-589A was detected up to 20~TeV. Its energy spectrum is well described by a power-law model, with a photon index $\Gamma$ = 2.2 and a mean differential flux  $N_{0}$ = (2.9$\pm$0.4)$\times 10^{-13}$ ph cm$^{-2}$ s$^{-1}$ TeV$^{-1}$ at 1~TeV. As in the case of other $\gamma$-ray binaries, the VHE spectrum cannot be extrapolated from the HE spectrum, which has a break  at around 1~GeV. The orbital light curve at VHE peaks in phase with the  X-ray and HE (1-10~GeV) light curves, see Fig.~\ref{fig:1fgl}.

%%%%%%%%%%%%%%%%%%%%%%%%%%%%%%%%%%%%%
%%%%%%%%%%%%%%%%%%%%%%%%%%%%%%%%%%%%% 

Based on optical spectroscopic observations, \cite{Strader+15} found that the maxima of the X-ray, HE, and VHE flux correspond to the inferior conjunction.  This finding was unexpected because $\gamma$-rays are believed to be produced through anisotropic IC up-scattering of the stellar UV photons. Therefore, the peak of the $\gamma$-ray flux should occur at the superior conjunction,  especially if the system is edge-on. This discrepancy could only be explained if the binary orbit is eccentric and the flux maximum occurs at periastron.

NuSTAR observations \cite{An+15} demonstrated that similar to other $\gamma$-ray binaries, the broad-band X-ray spectrum is well fitted with an unbroken power-law model. 
The source flux shows a correlation with the spectral hardness throughout all orbital phases.

A comparison of the light curves of \fgl\ at different energy ranges shows that both the X-ray and the low-energy ($E < 0.4$~GeV) $\gamma$-ray bands are characterised by a similar modulation (a broad maximum at $\phi = 0.2-0.7$ and a sharp spike at $\phi = 0$, see Fig.~\ref{fig:1fgl}), thus suggesting that they are due to a common spectral component. On the other hand, above $\approx 1$~GeV, the orbital light curve changes significantly because the broad hump disappears and the remaining structure is similar to the light curve observed at VHE. Based on these results, \cite{AnRomani17} suggested that the flux in the GeV band is due mainly to the pulsar magnetosphere, while the X-ray flux is due to synchrotron emission from shock-accelerated electrons and the TeV light curve is dominated by the up-scattering of the stellar and synchrotron photons through external Compton (EC) and synchrotron self-Compton (SSC) mechanisms, in an intrabinary shock. %In this frame, the 
The light curves at different energy ranges can be reproduced with the beamed SSC radiation from adiabatically accelerated plasma in the shocked pulsar wind. 
This is composed of a slow and a fast outflow. Both components contribute to the synchrotron emission observed from the X-ray to the low-energy $\gamma$-ray band, which has a sinusoidal modulation with a broad peak around the orbit periastron at $\phi$ = 0.4. On the other hand, only the Doppler-boosted component reaches energies above 1 GeV, which are characterised by the sharp maximum that occurs at the inferior conjunction at $\phi$ = 0.
This result can be obtained with an orbital inclination of $\approx$ 50$^{\circ}$ and an orbital eccentricity of $\approx$ 0.35, consistent with the constraints obtained from optical observations. In this way, the model could also explain the variable X-ray spike coincident with the $\gamma$-ray maximum at $\phi$ = 0.

\subsection{\hess0632}
In contrast to other $\gamma$-ray loud binaries, %\hess0632 
\object{HESS J0632+057} 
 remained the only system that for a time was lacking detection in the GeV energy band. Only recently have indications of a GeV detection with \fermi-LAT been reported by \cite{malyshev16} and \cite{li17}. The system was initially discovered during H.E.S.S.\ observations of the Monoceros region~\cite{hess_j0632} as an unidentified point-like source. Its spatial coincidence with the Be star MWC~148 suggested its binary nature~\cite{hess_j0632,hinton09}. 
% with unknown nature of the compact companion.
With dedicated observational campaigns, the binary nature of the system was confirmed by radio~\cite{skilton09} and soft X-ray~\cite{falcone10} observations. In the TeV band, the system was also detected by VERITAS and MAGIC~\cite{magic12,Aliu2014HESS0632}.

The orbital period of \hess0632 of $\approx 316 \pm 2$~d~\cite{malyshev17, j0632_icrc}, with a zero-phase time $T_0 = 54857$~MJD~\cite{bongiorno11}, was derived from \swift/XRT observations. The exact orbital solution and even the orbital phase of periastron is not firmly established and is placed at orbital phases $\phi\approx 0.97$~\cite{casares12} or $\phi \approx 0.4-0.5$~\cite{moritani18,malyshev17}.

\begin{figure}[t!]
\begin{center}
\includegraphics[width=0.27\linewidth]{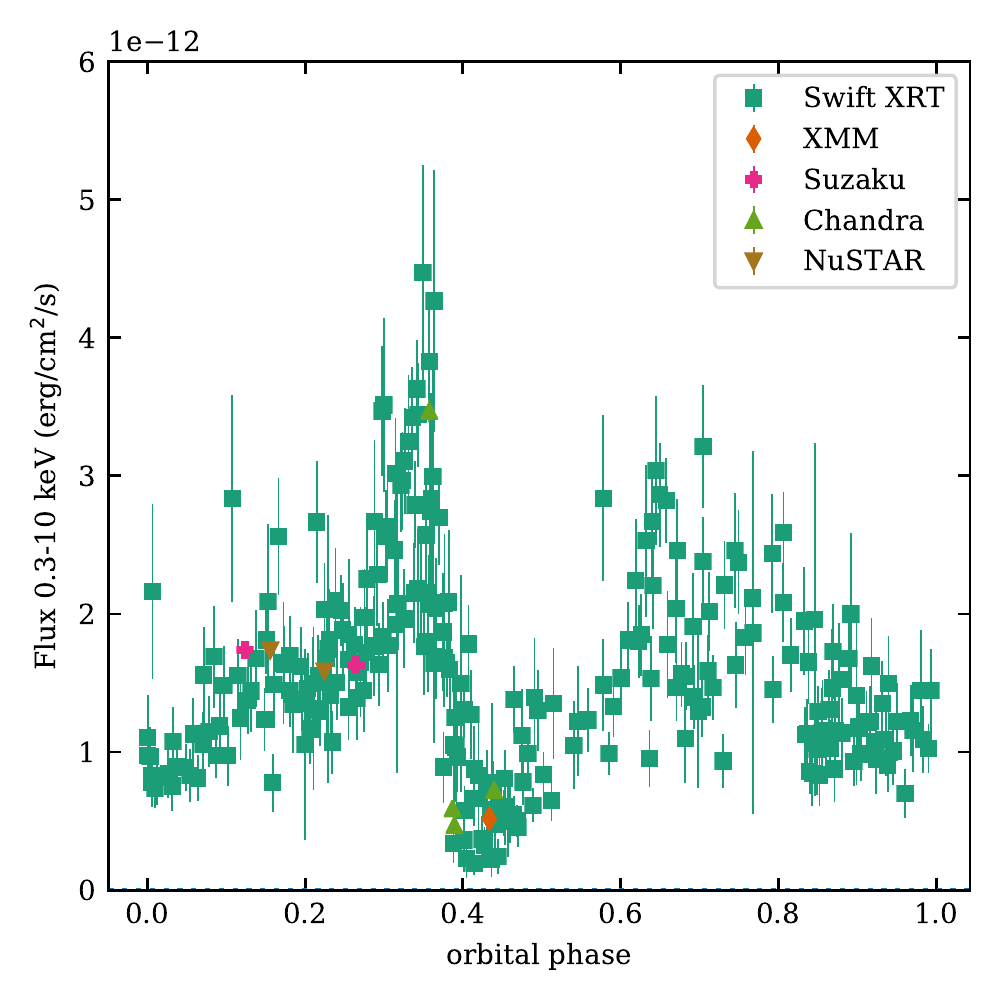}
\includegraphics[width=0.27\linewidth]{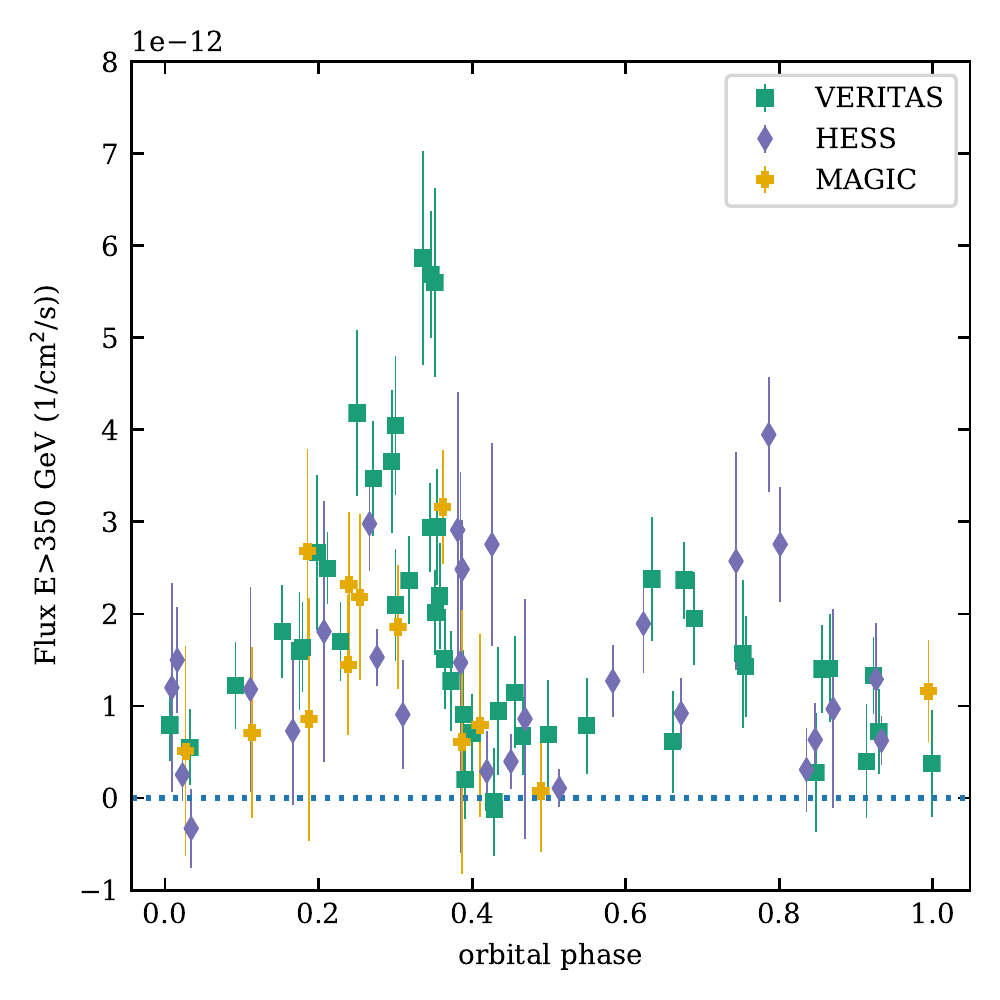}
\includegraphics[width=0.35\linewidth]{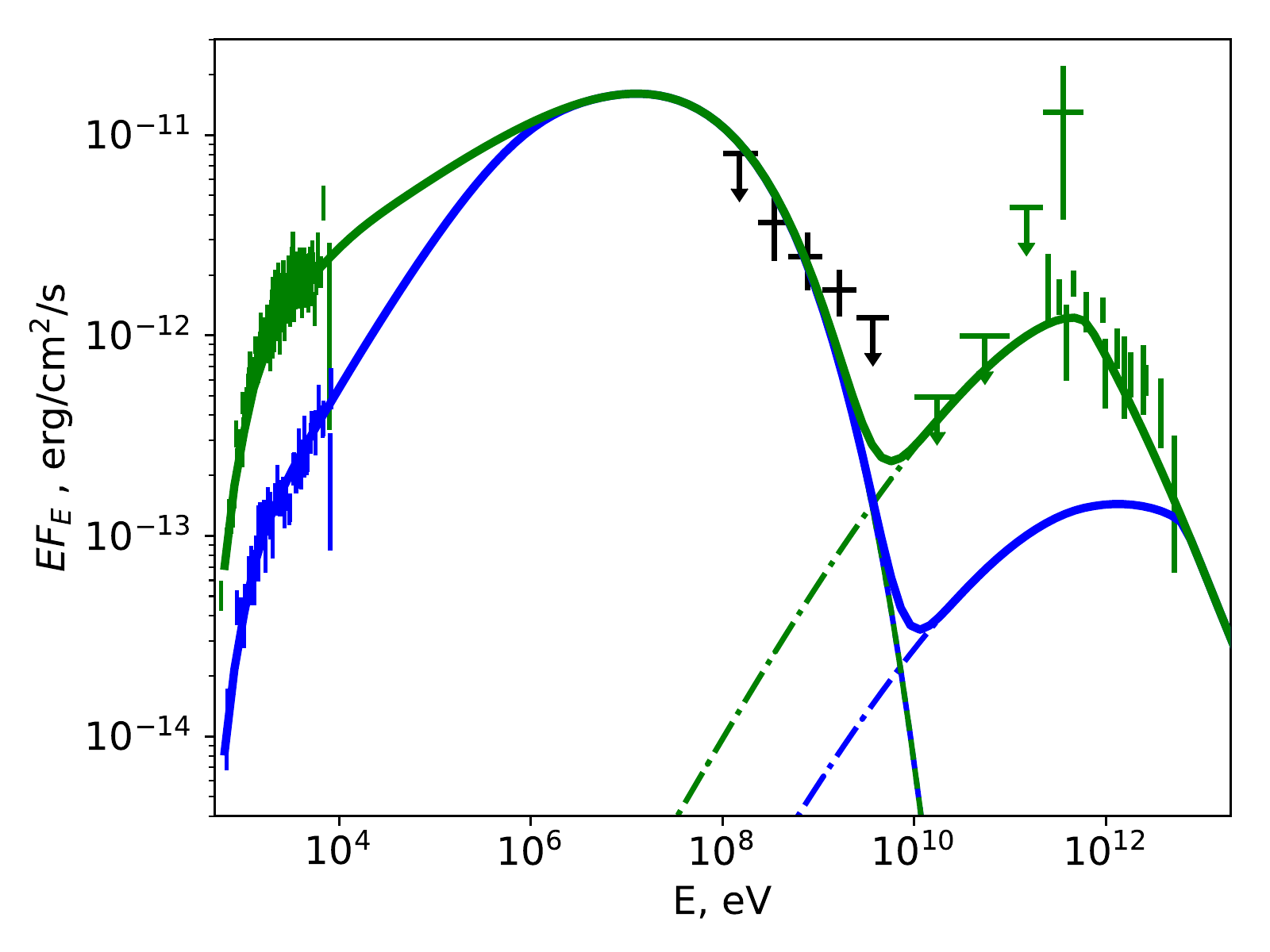}
\end{center}
\vspace*{-0.75cm}
\caption{ X-ray-to-TeV lightcurves (left and middle panels, adopted from~\cite{j0632_icrc}) and spectrum of \hess0632 during its high state (green points; orbital phases $\phi\approx 0.3-0.4$) and low state (blue points; $\phi \approx 0.4-0.5$). The data are adopted from~\cite{malyshev17} (X-rays), ~\cite{li17} (mean GeV spectrum, black points), and~\cite{malyshev16} (green upper limits). TeV data are adopted from~\cite{veritas15}. The solid lines show the ``similar to \psrb'' model flux, while dashed and dot-dashed lines illustrate contributions from synchrotron and IC model components correspondingly. See text for more details.}
\label{fig:model}
\end{figure}

The orbital folded X-ray light curve of \hess0632 has two clear emission peaks: first at phase $\phi\approx 0.2-0.4,$ and second at $\phi\approx 0.6-0.8$ separated by a deep minimum at $\phi\approx 0.4-0.5$~\cite{bongiorno11,Aliu2014HESS0632}, see Fig.~\ref{fig:model}. A low-intermediate state is present at $\phi\approx 0.8-0.2$. The orbital light curve in the TeV energy range shows a similar structure, as was reported by~\cite{veritas15}. Indications of orbital variability in the GeV range were reported by~\cite{li17}.

The X-ray-to-TeV spectrum of \hess0632 is shown in Fig.~\ref{fig:model}. Several models have been proposed so far to explain the observed variations of the flux and spectrum throughout the orbit. In the flip-flop scenario (see e.g. \cite{moritani15} and references therein) the compact object is assumed to be a pulsar that passes periastron at $\phi=0.97$. Close to apastron (orbital phases $\approx 0.4-0.6$), the pulsar is in a rotationally powered regime, while it switches into a propeller regime when periastron is approached (phases $0.1-0.4$ and $0.6-0.85$). When the gas pressure of the Be disc overcomes the pulsar-wind ram pressure, the pulsar wind in a flip-flop scenario is quenched (phases $0-0.1$ and $0.85-1$). Because the Be disc of the system is estimated to be about three times larger than the binary separation at periastron, the compact object enters a dense region of the disc near periastron. In this situation, the strong gas pressure is likely to quench the pulsar wind and suppress high-energy emissions.
%%%%%%%%%%%%%%%%%%%%%%%%%%%%%%%%%%%%%
%%%%%%%%%%%%%%%%%%%%%%%%%%%%%%%%%%%%%
Alternatively, the observed orbital variations can be explained within the ``similar to \psrb'' model~\cite{malyshev17}. The similar two-peak behaviour of the \hess0632 and \psrb orbital light curves allows us to assume that the orbital plane of \hess0632 is inclined with respect to the disc plane, similarly to \psrb. Orbital X-ray and TeV peaks within this model correspond to the first and second crossing of the disc by a compact object. Higher ambient density during these episodes leads to more effective cooling of the relativistic electrons by synchrotron and IC mechanisms, resulting in an increased level of X-ray and TeV emission. The orbital phase of periastron in this model is located at phase $\phi\approx 0.4 - 0.5$~\cite{malyshev17}.

The break in the GeV-TeV spectrum at $\approx 200$~GeV can be interpreted as a corresponding break in the spectrum of emitting relativistic electrons. The X-ray-to-GeV and TeV parts of the spectrum are explained as synchrotron and IC components. An initial power-law ($\Gamma_{1,e}\approx 1.3$) spectrum of electrons can be modified by synchrotron energy losses at above $E_{br}\approx 1$~TeV, resulting in a $\Gamma_{2,e}\approx 2.3$ higher energy slope. The absence of cooling in the energy band below 1~TeV could be attributed to the escape of the sub-TeV electrons from the system. A similar interpretation of the spectral energy distribution was proposed by~\cite{chernyakova15} for \psrb.

Alternatively, the spectral break in the electron spectrum can occur at the transition between the domination of adiabatic and IC or synchrotron losses 
(see e.g. \cite{khan07} and \cite{Suzaku2009} for \psrb and LS~5039). The adiabatic loss time is naturally shortest in sparse regions outside of the Be star disc and longest in dense regions inside it.

A broken power-law shape of the spectrum is not unique for the ``similar to \psrb'' model. A similar shape of the spectrum can also be expected within the flip-flop model because both interpretations of the break origin can be valid for this model. The two models can be distinguished by CTA observations of the variation in slope and low-energy break position throughout the orbit. 

Within the flip-flop model at orbital phases $\phi=0-0.4,$ the compact object moves from a denser to increasingly sparser regions of the Be star disc. The spectrum of relativistic electrons becomes increasingly less dominated by the losses. This results in a gradual hardening of the TeV slope and in a shift of the break energy to higher values. At phases $\phi\approx 0.6-1,$ the compact object enters denser regions of the disc, which should lead to a gradual softening of the slope and shift of the energy break to lower energies. The spectrum is expected to be hardest when the object is beyond the Be star disc (orbital phase $\approx 0.4$). This phase corresponds to the minima of observed emission. The softest spectrum is expected when the compact object approaches periastron, that is, at phase $\phi\approx 0.97$.

In the ``similar to \psrb'' model the compact object intersects the disc of the Be star twice per orbit (at orbital phases $0.2-0.4$ and $0.6-0.8$) where the soft spectrum with the low position of energy break is expected. At phases $0-0.2$, $0.4-0.6$ and $0.8-1$ in the ``similar to \psrb'' model, the compact object is beyond the dense regions of the disc. At these orbital phases a hard slope with energy break shifted to higher energies can be expected.

\section{Poorly studied sources}
\label{sec:poor}
In this section we briefly discuss sources which either do not have unambiguous binary interpretation (\hess1832), or recently discovered binary systems on which little data is available(\lmc and 4FGL J1405.1-6119) 

\subsection{\hess1832}
\object{HESS J1832-093}
is a   $\gamma$-ray binary candidate discovered  as a TeV point source by H.E.S.S. This source lies in the vicinity of SNR G22.7-0.2, which can suggest its possible association with this SNR \cite{abramowski15}. However, several follow-up observations in X-rays  instead support the binary nature of this source~\cite{eger16,mori17}. A simple power law model well describes the TeV  spectrum with a photon index of $\Gamma$ = 2.6 $\pm$ 0.3 and an integrated photon flux above 1 TeV of $F = (3.0 \pm 0.8 )\times 10^{-13}$~cm$^{-2}$~s$^{-1}$~\cite{abramowski15}. An {\it XMM-Newton} observation of the source field discovered a bright X-ray source, XMMU J183245-0921539 within the $\gamma$-ray error circle~\cite{abramowski15}. This source is also associated with a point source detected in a subsequent \cha observation campaign~\cite{eger16}. During the \cha observations, an increase of the 2-10 keV flux of the order of 4 with respect to the earlier \xmm measurement and the coincidence of a bright IR source at the \cha error box suggest a binary scenario for the $\gamma$-ray emission~\cite{eger16}.  

4FGL J1832.9-0913 is a $\gamma$-ray source included in the 4FGL catalog,  close  to  the  position  of  HESS J1832-093  and  spatially compatible with the SNR G22.7-0.2. Intensive Swift monitoring 
of the system allowed \cite{2020arXiv200102701M} to find a period of about 86 days both in X-ray and GeV energy bands. The overall spectral 
energy distribution of the source strongly resembles the known  $\gamma$-ray binary HESS J0632+057 and peaks in $\gamma$-rays.  The TeV and GeV spectral components do not arise from a single power law, since the upper limits at intermediate $\gamma$-rays prevent such connection. This is similar to what has been observed
in some other $\gamma$-ray binaries, like \lsi, \ls and HESS J0632+057. 
Near infrared spectroscopy of the IR counterpart of HESS J1832-093, allowed \cite{tam2020} 
to identify  it as an O-, or less likely, early B-type star.

\subsection{\lmc}

\object{LMC P3} is the first and, up to the moment, the only known extragalactic \grb. It was detected in 2016 with the \flat in the Large Magellanic Cloud from a search for periodic modulation in all sources in the third \flat catalog \cite{3FGL}. The system has an orbital period of 10.3 days and is associated with a massive O5III star located in the supernova remnant DEM L241 \cite{2016ApJ...829..105C}. \swift/XRT X-ray  and ATCA radio observations demonstrated that both X-ray and radio emission are also modulated on the 10.3 day period, but are in anti-phase with the $\gamma$-ray modulation. The X-ray spectrum is well described by a single power law with $\Gamma=1.3\pm0.3$, modified by a fully covered absorber. The resulting value of the hydrogen column density of a fully covered absorber is comparable with the Galactic HI value.

Optical radial velocity measurements suggest that, unless the system has a very low inclination the system contains a NS \cite{2016ApJ...829..105C}. Low inclinations, however, result in a range  of masses of the compact object above the Chandrasekhar limit, e.g. a BH with a mass of  M=5M$_\odot$ will have an inclination $i=14{^{+4}_{-3}}^\circ$, and   $i=8\pm 2^\circ$  for M=10M$_\odot$.  The source is significantly more luminous than similar sources in the Milky Way at radio, optical, X-ray and $\gamma$-ray wavelengths. It is at least four times more luminous in GeV gamma rays and 10 times more luminous in radio and X-rays than LS 5039 and 1FGL~J1018.6-5856, though the luminosity of the companion star and the orbital separations are comparable in all three systems.

The LMC has been observed extensively with H.E.S.S. since 2004. The data which were collected for the LMC between 2004 and the beginning of 2016 results in an effective exposure time for LMC P3 of 100 hours \cite{2018arXiv180106322H}. The sensitivity of H.E.S.S. does not allow a detection of flux variations of the object on a nightly basis. 
The low flux coming from the system does not allow for any statistically significant detection of periodicity using a Lomb-Scargle test and the Z-Transformed Discrete Correlation Function.
Folding the light curve with the orbital period of the system of 10.301 days, clearly demonstrates the orbital modulation of the VHE with a significant detection only in the orbital phase bin between 0.2 and 0.4 (orbital phase zero is defined as the maximum of the HE light curve at MJD 57410.25). The H.E.S.S. spectrum during the on-peak part of the orbit is described by a power-law with a photon index $\Gamma=2.1\pm0.2$. The averaged slope along the total orbit is softer with $\Gamma=2.5\pm0.2$. The VHE flux above 1 TeV varies by a factor more than 5 between on-peak and off-peak parts of the orbit.

The minimum HE emission occurs between orbital phases 0.3 - 0.7. 
The shift between the orbital phase of HE and VHE peaks is not unique to this $\gamma$-ray binary.
For example, 
a similar shift
is observed in \ls (see Section \ref{section:ls}), as the angle-dependent cross section of IC scattering and $\gamma \gamma$ absorption due to pair-production affects the HE and VHE in  different ways~(e.g. \cite{LS5039_2008_Dubus,LS5039_2008_Khan,2008ApJ...672L.123N}).

Recently reported optical spectroscopic observations of \lmc\ have better constrained the orbital parameters \cite{2019MNRAS.484.4347V}. The observations find the binary has an eccentricity of $0.4\pm0.07$ and place superior conjunction at phase $\sim0.98$ and inferior conjunction at phase $\sim 0.24$. These phases correspond to the points of the maxima reported in \flat and H.E.S.S.\ light curves respectively. The mass function found ($\sim 0.0010 $ M$_\odot$) favours a NS companion, for most inclination angles.

The detection of VHE emission during the entire orbit is critical for detailed modelling that will allow us to understand what is happening in the system. 

\subsection{4FGL J1405.1-6119}
\object{4FGL J1405.1-6119} has been identified as a gamma-ray binary from the modulation of gamma-ray flux, which shows a period of
13.7 days \cite{Corbet19_4fgl}.  Gamma-ray emission is characterised with the presence of two
maxima per orbit.
SWIFT/XRT observations show that X-ray
emission is also modulated at this period, but with a single maximum that is closer to the secondary
lower-energy gamma-ray maximum. A radio source, coincident with the X-ray source is also found
from Australia Telescope Compact Array  observations, and the radio emission is modulated
on the gamma-ray period with similar phasing to the X-ray emission. As with 1FGL J1018.6-5856 and LMC
P3, which were also detected from modulated gamma-ray
emission with the LAT, 4FGL J1405.1-6119 contains an
O, rather than a Be, star primary.  For a distance of 7.7 kpc, the implied maximum
gamma-ray luminosity is comparable to, and
possibly higher than, that of LS 5039, while approximately half that of 1FGL J1018.6-5856 and a tenth that of LMC P3. The system is heavily
obscured in the optical band. Further multiwavelength observations are needed to better constrain physical properties in the system.

%\subsection{HESS J1832-093}

\section{Microquasars}
\label{sec:microq}
Microquasars are binary systems with a compact object (NS or a BH) orbiting around an optical star. Matter outflow from the optical star lead to formation of an accretion disk around the compact object and a relativistic collimated jet. These binary star systems are known under the name of `microquasars' 
because  the physical processes taking place in these systems mimic on a smaller scales ones happening in quasars (`quasi-stellar-radio-source'). The first time this term was used in 1992 following the discovery of the relativistic radio jet from  1E1740.7-2942,  one of the most  luminous  sources  of  soft  gamma-rays  in  the  Galactic  Center  region \cite{Mirabel92}.

At the moment more than 20 microquasars are known. Subsequent observations demonstrated correlation between the mass of the compact object,  radio (5 GHz) and X-ray (2–10 keV) luminosities (e.g. \cite{Falcke2004}), strengthening the link between AGNs and microquasars.  In AGNs, jets are known to be places of efficient particle acceleration  and produce broad band non-thermal photon emission. The resulting radiation emission can extend from radio up to the very high-energy (VHE; E>100 GeV) band. According to the TeVCat \footnote{http://tevcat2.uchicago.edu/} more than 65 AGNs has been already detected by current Cherenkov telescopes. If similar jet production and  particle acceleration mechanisms operate in microquasars and AGNs, this might imply that microqusars should be
sources of  VHE $\gamma$-ray emission as well. At the moment only 3 microqusars, Cyg X-1, Cyg X-3 and SS433, are detected at gamma-rays.

%%%%%%%%%%%%%%%%%%%%%%%%%%%%%%%%%%%%%
\begin{figure}[t!]
\vspace*{-0.5cm}
\begin{center}
\includegraphics[angle=0,width=0.31\linewidth]{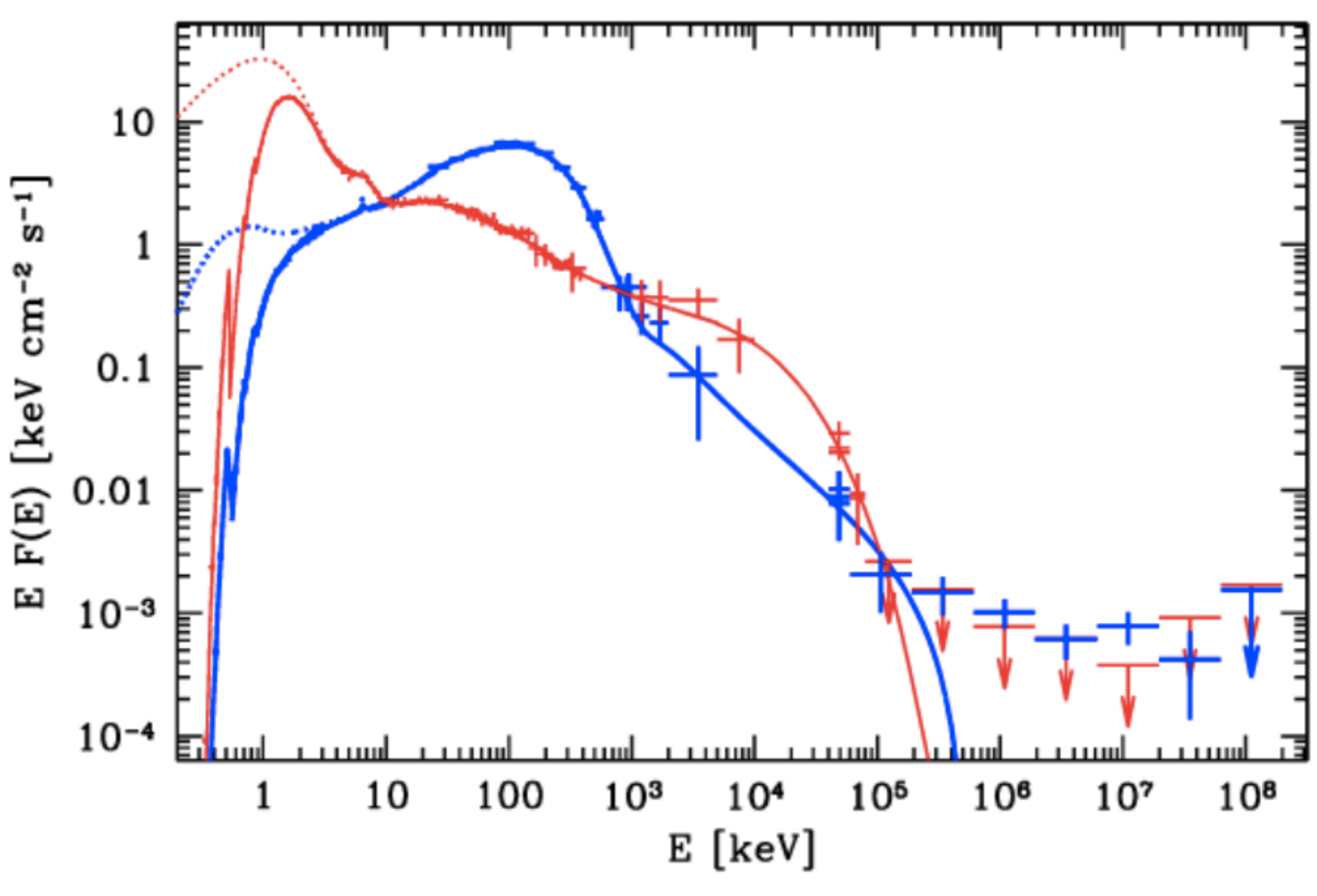}
\includegraphics[angle=0,width=0.31\linewidth]{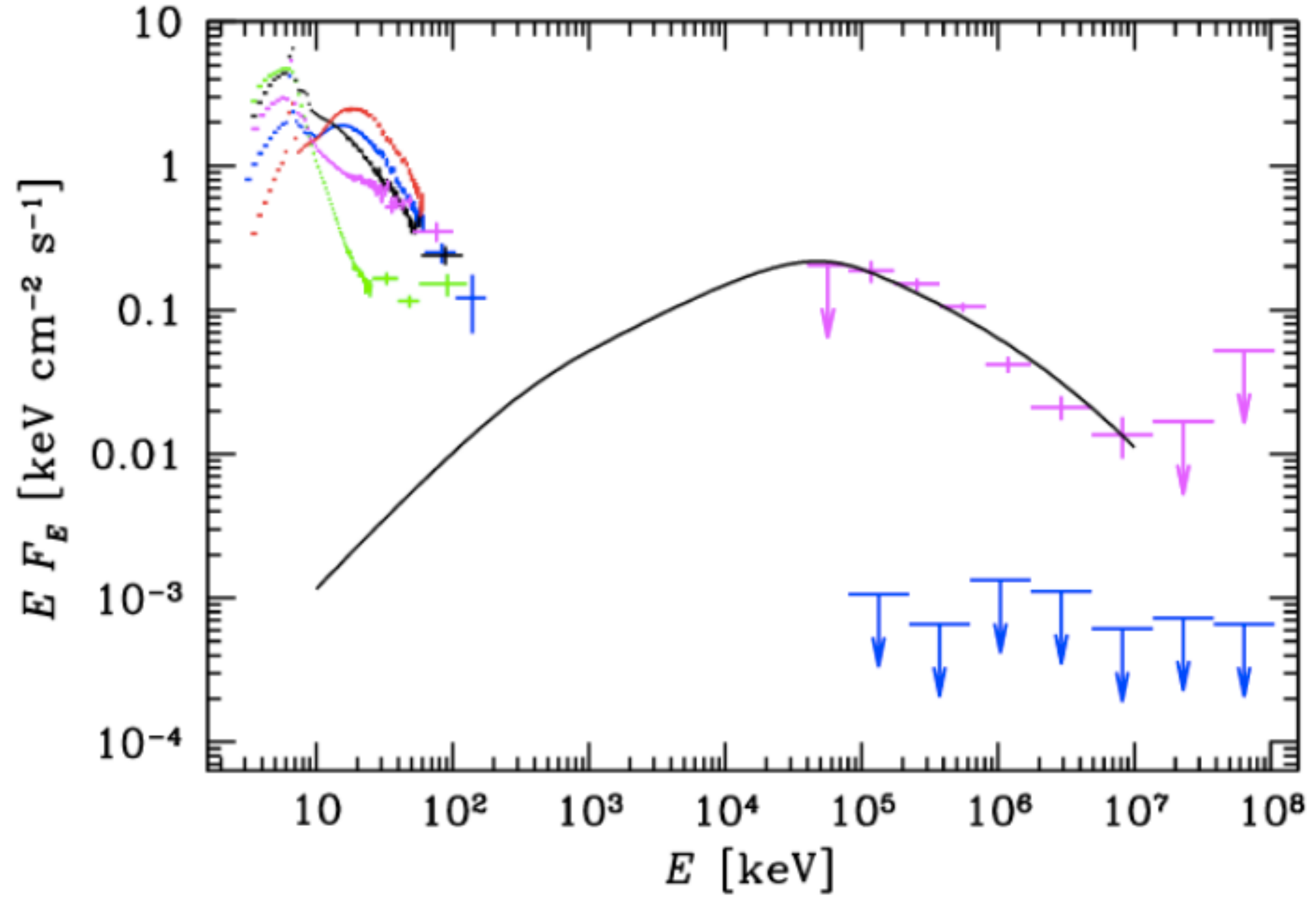}
\includegraphics[angle=0,width=0.33\linewidth]{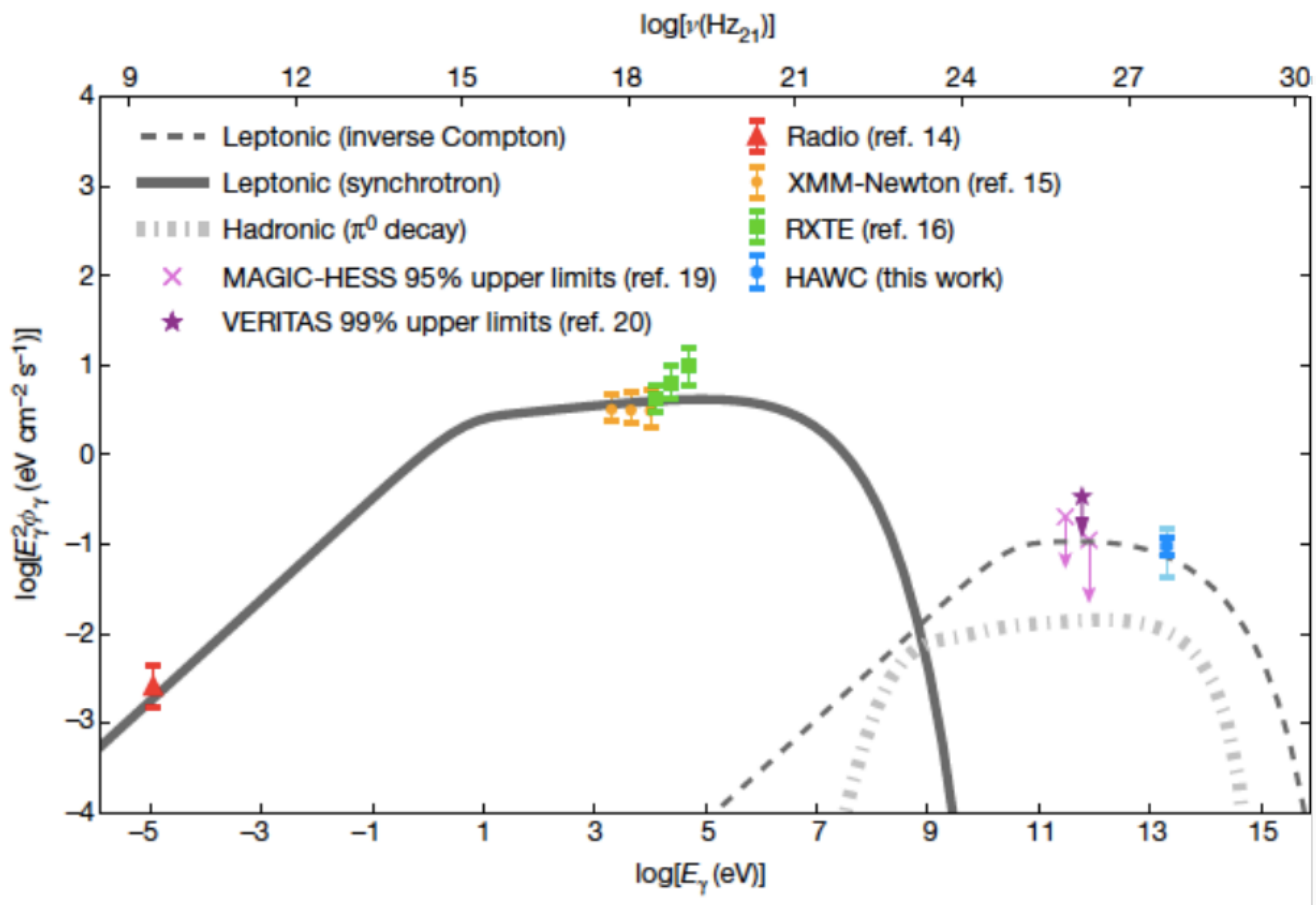}
\end{center}
\vspace*{-0.75cm}
\caption{\textit{left:} Broad-band spectra of Cyg X-1 in the hard (blue) and soft (red) states compared to hybrid-Comptonization accretion-flow models, adapted from \cite{CygX1_2017_Zdz}. \textit{middle:}  Broad-band spectra of Cyg X-3 in the soft (magenta)  and hard (blue) states,  adapted from \cite{2018MNRAS.479.4399Z}). \textit{right:}Broadband spectral energy distribution of the eastern emission region e1 of SS433, adapted from \cite{SS433_HAWC}. }
\label{fig:SS433}
\end{figure}
%%%%%%%%%%%%%%%%%%%%%%%%%%%%%%%%%%%%%

\subsection{Cyg X-1}

\object{Cyg X-1} is an X-ray binary with  a (14.8$\pm$1.0) $M_{\bigodot}$ BH on an almost circular 5.6 d orbit around a (19.2$\pm$1.9) M$_{Sun}$ 
O9.7Iab supergiant star \cite{1999MNRAS.309.1063B}. Orbital period is detected in both  X-ray and radio energy bands due to  absorption/scattering of the radiation by the stellar wind \cite{2002MNRAS.336..699B,2007MNRAS.375..793S}. In X-rays source also exhibit a super-orbital variability with a 300-day  period \cite{2011MNRAS.412.1985Z}. Cygnus X-1 was classified as a microquasar after the detection of a one-sided relativistic radio-jet \cite{2001MNRAS.327.1273S}. It displays the two principal X-ray states of BH transients, the soft state  and the hard state (see left panel of Figure \ref{fig:SS433}). Both are described by the sum of a blackbody-like emission from the accretion disk that peaks at 1 keV (dominant in the soft state) and a power-law tail (dominant in the hard state). During hard state  the source displays persistent jets from which synchrotron radio emission is detected, whilst in the soft state, these jets are disrupted.

 In the hard state, the main component of the X-ray spectrum is thermal Comptonization in a plasma with the electron temperature of about 100 keV, which features a sharp cut-off at about 200 keV. Beyond 1 MeV, there is a clear high-energy tail, measured upto 3 MeV. The origin of the photon tail may be Compton scattering by a power-law electron tail above the thermal electron distribution in the accretion flow. In the soft state, there is a strong disc black body component in the X-ray spectrum, peaking at 1 keV, followed by a high-energy tail, measured up to 10 MeV (e.g. \cite{McConnell2002}).

Analysis of the \flat data demonstrated the presence of the  steady high-energy $\gamma$-ray emission  from Cyg X-1 in the hard and intermediate spectral states \cite{2013MNRAS.434.2380M,CygX1-2016-Zanin,CygX1_2017_Zdz}. \cite{CygX1_2017_Zdz} explained the observed hard-state $\gamma$-ray emission above 100 GeV spectra within a self-consistent jet model, taking into account clumping and all the relevant emission and absorption processes.  In the soft spectral state, the emission is detected only up to 80 MeV. Detected emission below 100 MeV is  well explained by the high-energy tails of the emission of the accretion flow, in both the hard and the soft states \cite{CygX1_2017_Zdz}.

First Magic observations of Cyg X-1 in 2006 resulted in a tentative detection of a flare from the source  at a 3.2 $\sigma$ confidence level \cite{CygX1Magic07}. This detection coincided with an X-ray flare seen by RXTE, Swift and INTEGRAL observatories.  After the first detection in 2006 MAGIC performed  observations of the source for 100 hours between 2007 and 2014 \cite{2017MNRAS.472.3474A}, mainly during hard state. No significant excess above 200 GeV was detected for steady, orbital or daily basis emission. An orbital-phase analysis did not show any orbital modulation neither during hard (83 hours of observation) neither during soft (14 hours) states.  These results rule out VHE emission from the large scale jet,  or from the interaction between the jet and surrounding medium above the sensitivity level of MAGIC. More sensitive instruments as CTA would be required to detect steady TeV emission.

\subsection{Cyg X-3}

Cyg X-3, one of the first discovered X-ray binaries \cite{1967ApJ...148L.119G}, is a unique and puzzling system. Cyg X-3 is the only known binary in the Galaxy containing both a compact object and a Wolf-Rayet star  with a very short orbital period of 4.8 h. The compactness of the system produces an unusually high absorption, probably caused by the wind of the companion star, which complicates the identification of the compact object (1.4 M$_{Sun}$ NS, or less than 10 M$_{Sun}$ BH). Despite  this  strong  X-ray  absorption,  the X-ray spectrum  shows two main spectral X-ray states resembling the canonical states of the BH binaries (see middle panel of Figure \ref{fig:SS433}). 
The hard state is characterized by a weak soft thermal component and a strong non-thermal power-law emission that peaks at hard X-ray energies, whereas the soft state, though showing a non-thermal tail \cite{2008MNRAS.388.1001S}, is dominated by the optically thick thermal disk emission. In Cygnus X-3, however, the hard state displays a high-energy cutoff at $\sim$ 20 keV, significantly lower than the 100 keV value found for BH binaries. 

Cygnus X-3 is the strongest radio source among the X-ray binaries, whose flux can vary several orders of magnitude during its frequent radio outbursts. These major flares happen only in soft state \cite{2008MNRAS.388.1001S}. High-energy $\gamma$-ray emission from the system was discovered by the \flat  and AGILE \cite{2009Sci...326.1512F,2009Natur.462..620T} in the soft spectral state.  The GeV power-law emission and its orbital modulation appear to be due to Compton upscattering of the stellar emission from the companion WR star by jet relativistic electrons with a power-law distribution with the spectral index of 3.5 and the low-energy cutoff at the Lorentz factor of 10$^3$ \cite{2010MNRAS.404L..55D,2018MNRAS.479.4399Z}. Along with Cyg X-1,  Cyg X-3 is one of only two X-ray binaries that are certainly powered by accretion for which non-extended HE $\gamma$-ray emission  has been detected at a high statistical significance.   

Despite observing the source during strong radio and HE outbursts, no significant excess was found  by  MAGIC.  One  has  to  consider  the  extremely  high  absorption  due  to  the  Wolf-Rayet,  which may affect VHE gamma-ray emission. Note however a report from SHALON mirror Cherenkov telescope \cite{2018AstL...44..162S}, where authors claim a detection of a  VHE $\gamma$-ray source, positionally  coincident with the microquasar Cyg X-3, and revealing variability on 4.8 h  time scale.

\subsection{SS 433}
SS 433 is a binary system containing a supergiant star that is overflowing its Roche lobe with matter accreting onto a compact object (either a BH or a NS) (e.g. \cite{1984ARA&A..22..507M,2004ASPRv..12....1F}). Two jets of ionised matter with a bulk velocity of approximately 0.26c (where c is the speed of light in vacuum) extend from the binary, perpendicular to the line of sight, and terminate inside  a supernova remnant W50 (e.g.\cite{2004ASPRv..12....1F}). The lobes of W50 in which the jets
terminate, about 40 parsecs from the central source, are  accelerating charged particles, as follows from  radio and X-ray observations, consistent with electron synchrotron emission \cite{1980A&A....84..237G,2007A&A...463..611B}.

\cite{SS433_HAWC} reported results of High Altitude Water Cherenkov (HAWC) Observatory observations of SS 433  collected between November 2014 and  December 2017. In 1017 days of measurements with HAWC, an excess of $\gamma$-rays with a post-trials significance of 5.4 $\sigma$ has been observed in a joint fit of the eastern and western interaction regions of the jets of SS 433. The $\gamma$-rays detected by HAWC can be  produced either by protons with energy of at least 250 TeV, or by
electrons of at least 130 TeV to up-scatter the cosmic microwave background (CMB) low-energy photons to 25-TeV $\gamma$-rays. 
%In this case, the radio to X-ray emission is dominated by synchrotron radiation from the same population of electrons in the magnetized plasma of the jets and lobes.

\cite{SS433_HAWC} modelled  broadband spectral energy distribution of the eastern emission
region e1 (see right panel of Figure \ref{fig:SS433}) and showed that the flux of VHE $\gamma$- rays observed by HAWC makes the proton scenario for SS 433 unlikely, because the total energy required to produce the highly relativistic protons is too high, as  the jets of SS 433 are known
to be radiatively inefficient, with most of the jet energy transformed
into the thermal energy of W50 \cite{2017A&A...599A..77P} rather than into particle acceleration.

Highly relativistic electrons, on the other hand, can produce $\gamma$-rays much more efficiently, primarily via IC scattering of CMB photons to $\gamma$-rays. The IC losses due to upscattering of infrared and optical photons are suppressed owing to the Klein–Nishina effect and are thus dominated by scattering of CMB photons \cite{2005MNRAS.363..954M}. The observation from HAWC suggests that the highly energetic electrons in SS 433 are probably accelerated in the jets and near the VHE $\gamma$-ray emission regions \cite{SS433_HAWC}.

Analysis of  the Fermi LAT  data,
led to the discovery of the  significant $\gamma$-ray emission from the region around SS433 \cite{SS433_bordas15,SS433_extFermi19,SS433_varFermi19,SS433_Fermi19}.
The analysis are very model dependent and can lead to very different conclusions. In \cite{SS433_Fermi19} authors use 10 years of  the Fermi LAT  data, and conclude that a stable emission is coming from the western interaction region in W50 at a level greatly exceeding predictions of the leptonic model discussed in \cite{SS433_HAWC}.  In \cite{SS433_varFermi19} authors are using 9 years of Fermi-LAT data, and third Fermi catalogue (3FGL), which includes updates of
the Galactic and extragalactic diffuse models, not used in \cite{SS433_Fermi19}. Usage of the 3FGL catalogue lead authors for a conclusion that the observed GeV emission is centred on SS43.  Authors also note that the GeV emission is probably extended with a best-fit extension of (0.84$\pm$0.27)$^\circ$. The
spectrum is best fit by a log-parabola with no evidence for emission above 500MeV.
In \cite{SS433_varFermi19} authors report an  evidence at the  3 $\sigma$ level for modulation of the $\gamma$-ray emission with the precession period of the jet, but no significant evidence for orbital modulation of the emission.
These results suggest that at least some of SS433's $\gamma$-ray emission originates close to the base of the jet. In \cite{SS433_extFermi19} authors analyze 10 years of Fermi-LAT data towards the SS433/W50 region. Using the latest source catalog and diffuse background models, they conclude that  SS433/W50 is detected with a significance of  $\sim 6\sigma$ in the photon energy range of 500 MeV - 10 GeV. In this energy range an extended flat disk morphology with radius of (0.45$\pm$0.06)$^\circ$ shifted by and 0.2$^\circ$ away from the nominal position  is preferred over a point-source description, suggesting that the GeV emission
region is much larger than that of the TeV emission detected by HAWC. The size of the GeV emission is instead consistent with the
extent of the radio nebula W50, so that the GeV emission may originate from this supernova remnant. Thus GeV results  are very dependent on the selection of the diffuse background models and dedicated analysis on the possible systematic errors is still needed.

\bigskip

\noindent\textbf{Acknowledgements}

The authors would like to acknowledge networking support by the COST Actions CA16214 and CA16104. We acknowledge SFI/HEA Irish Centre for High-End Computing (ICHEC) for the provision of computational facilities and support.

\bibliographystyle{aa}

% Bibliography and bibfile
\def\aj{AJ}%
          % Astronomical Journal
\def\actaa{Acta Astron.}%
          % Acta Astronomica
\def\araa{ARA\&A}%
          % Annual Review of Astron and Astrophys
\def\apj{ApJ}%
          % Astrophysical Journal
\def\apjl{ApJ}%
          % Astrophysical Journal, Letters
\def\apjs{ApJS}%
          % Astrophysical Journal, Supplement
\def\ao{Appl.~Opt.}%
          % Applied Optics
\def\apss{Ap\&SS}%
          % Astrophysics and Space Science
\def\aap{A\&A}%
          % Astronomy and Astrophysics
\def\aapr{A\&A~Rev.}%
          % Astronomy and Astrophysics Reviews
\def\aaps{A\&AS}%
          % Astronomy and Astrophysics, Supplement
\def\azh{AZh}%
          % Astronomicheskii Zhurnal
\def\baas{BAAS}%
          % Bulletin of the AAS
\def\bac{Bull. astr. Inst. Czechosl.}%
          % Bulletin of the Astronomical Institutes of Czechoslovakia
\def\caa{Chinese Astron. Astrophys.}%
          % Chinese Astronomy and Astrophysics
\def\cjaa{Chinese J. Astron. Astrophys.}%
          % Chinese Journal of Astronomy and Astrophysics
\def\icarus{Icarus}%
          % Icarus
\def\jcap{J. Cosmology Astropart. Phys.}%
          % Journal of Cosmology and Astroparticle Physics
\def\jrasc{JRASC}%
          % Journal of the RAS of Canada
\def\mnras{MNRAS}%
          % Monthly Notices of the RAS
\def\memras{MmRAS}%
          % Memoirs of the RAS
\def\na{New A}%
          % New Astronomy
\def\nar{New A Rev.}%
          % New Astronomy Review
\def\pasa{PASA}%
          % Publications of the Astron. Soc. of Australia
\def\pra{Phys.~Rev.~A}%
          % Physical Review A: General Physics
\def\prb{Phys.~Rev.~B}%
          % Physical Review B: Solid State
\def\prc{Phys.~Rev.~C}%
          % Physical Review C
\def\prd{Phys.~Rev.~D}%
          % Physical Review D
\def\pre{Phys.~Rev.~E}%
          % Physical Review E
\def\prl{Phys.~Rev.~Lett.}%
          % Physical Review Letters
\def\pasp{PASP}%
          % Publications of the ASP
\def\pasj{PASJ}%
          % Publications of the ASJ
\def\qjras{QJRAS}%
          % Quarterly Journal of the RAS
\def\rmxaa{Rev. Mexicana Astron. Astrofis.}%
          % Revista Mexicana de Astronomia y Astrofisica
\def\skytel{S\&T}%
          % Sky and Telescope
\def\solphys{Sol.~Phys.}%
          % Solar Physics
\def\sovast{Soviet~Ast.}%
          % Soviet Astronomy
\def\ssr{Space~Sci.~Rev.}%
          % Space Science Reviews
\def\zap{ZAp}%
          % Zeitschrift fuer Astrophysik
\def\nat{Nature}%
          % Nature
\def\iaucirc{IAU~Circ.}%
          % IAU Cirulars
\def\aplett{Astrophys.~Lett.}%
          % Astrophysics Letters
\def\apspr{Astrophys.~Space~Phys.~Res.}%
          % Astrophysics Space Physics Research
\def\bain{Bull.~Astron.~Inst.~Netherlands}%
          % Bulletin Astronomical Institute of the Netherlands
\def\fcp{Fund.~Cosmic~Phys.}%
          % Fundamental Cosmic Physics
\def\gca{Geochim.~Cosmochim.~Acta}%
          % Geochimica Cosmochimica Acta
\def\grl{Geophys.~Res.~Lett.}%
          % Geophysics Research Letters
\def\jcp{J.~Chem.~Phys.}%
          % Journal of Chemical Physics
\def\jgr{J.~Geophys.~Res.}%
          % Journal of Geophysics Research
\def\jqsrt{J.~Quant.~Spec.~Radiat.~Transf.}%
          % Journal of Quantitiative Spectroscopy and Radiative Trasfer
\def\memsai{Mem.~Soc.~Astron.~Italiana}%
          % Mem. Societa Astronomica Italiana
\def\nphysa{Nucl.~Phys.~A}%
          % Nuclear Physics A
\def\physrep{Phys.~Rep.}%
          % Physics Reports
\def\physscr{Phys.~Scr}%
          % Physica Scripta
\def\planss{Planet.~Space~Sci.}%
          % Planetary Space Science
\def\procspie{Proc.~SPIE}%
          % Proceedings of the SPIE
\let\astap=\aap
\let\apjlett=\apjl
\let\apjsupp=\apjs
\let\applopt=\ao
\bibliography{reference}

\end{document}